\documentclass[11pt,a4paper]{article}
\usepackage{jcappub}

\usepackage{lineno}

\title{Temporal Invariance Is an Illusion: Time-Dependent Influences of the Galactic Magnetic Field on UHECR Observations}

\author{Veronika Vašíčková,}
\author{Leonel Morejon,}
\author{Karl-Heinz Kampert}

\affiliation{Bergische Universität Wuppertal, Gaußstr. 20, 42119 Wuppertal, Germany}

\emailAdd{vasickova@uni-wuppertal.de}

\abstract{
Understanding the origin of the Ultra-High-Energy Cosmic Rays (UHECRs) requires explaining the features of their energy spectrum, mass composition, and arrival directions. Current modeling approaches neglect the time evolution of UHECR observables, a factor that is particularly important in the case of bursting UHECR sources. This study focuses on the influence of time delays caused by the galactic magnetic field (GMF) on the spectrum and arrival directions of UHECRs observed on Earth. Using CRPropa 3.2, we investigate the rigidity-dependence of the residence time of extragalactic cosmic rays entering our Galaxy. We find that UHECRs entering the Milky Way can experience delays of hundreds of kiloyears relative to light, and we demonstrate that these delays significantly alter the UHECR observables. Notably, a cutoff emerges in the transient scenario within the rigidity range of $10^{18}-10^{19}$~V, which coincides with the spectral break observed in data. We find a progressive shift in composition favoring heavier nuclei, as well as a delay distribution that is correlated with GMF strength. 
This causes the particles to be less correlated with their initial direction the larger their delays.
A dipole-like anisotropy develops over timescales of about $\sim$100 kyr in certain bursts scenarios. Our results provide an alternative explanation for the UHECR spectral cutoff that does not invoke limits on source acceleration. This could potentially revise existing constraints.
}

\begin{document}
\maketitle
\flushbottom


\section{Introduction}

Inferring information about the sources of ultrahigh-energy cosmic rays (UHECRs) requires understanding the effects of magnetic fields on their spectrum and arrival directions. As Galactic magnetic field (GMF) models have been improved, exploring their effects on UHECRs has become a primary motivation. However, despite recent improvements \citep[e.g.][]{Unger:2023lob,Korochkin:2024yit}, the recognized effects on UHECRs have not fundamentally changed from those described with simpler models \citep{Harari:1999it,Harari:2000az}. These effects amount to moderate and regular deviations in the arrival directions of UHECR, which can induce energy-dependent caustics analogous to the optical effects of lenses and are manifested by the appearance of multiple images and flux magnifications. These effects have been demonstrated through the modeling of cosmic rays with rigidities $R=E/Z$ (where $E$ and $Z$ are the energy and atomic number of the UHECR, respectively) ranging from a few to 100 EV and have been further investigated in other studies \citep{Giacinti:2011uj}. Combined fits of the UHECR spectrum and composition consistently yield the best results for maximum  source rigidities of $R_{\rm max}=$4-5\,EV \cite{PierreAuger:2016use,PierreAuger:2023htc}. This rigidity limit was found to be quite robust and has led to the hypothesis that the sources must be almost identical \cite[e.g.][]{Ehlert:2022jmy}. Below these rigidities, UHECRs deviate significantly from their initial arrival directions, particularly due to the dominant coherent components of the GMF. Detailed simulations performed in Ref.\,\cite{Farrar:2017lhm} using the JF12 \citep{Jansson:2012pc} model, discussed GMF effects for rigidities down to 1\,EV. While these detailed simulations only explored a few sources and values of incoming directions and propagation distances, they found that, below 10\,EV, the deflection angles increase rapidly. However, the impact of the incidence position in the Galaxy and the residence time were not examined.

Multiple recent studies \citep[e.g.][]{Bister:2024ocm} are based on the results and simulation data from \cite{Farrar:2017lhm}, which uses lensing techniques involving the backtracking of particles from Earth. This approach was first suggested in \citep{Thielheim:1968} and first applied in \citep{Harari:1999it}. 
Lensing techniques map observed directions on Earth to incoming directions outside the Galaxy as a function of rigidity. This method is computationally efficient because it only considers particles that arrive on Earth, however, it has several limitations. For instance, it disregards the residence time of extragalactic UHECR in the Galaxy, as discussed in Refs.\,\cite{Kaapa:2022tey,Kaapa:2022qqn}.

Previous estimates of the time delay accumulated by cosmic rays in the Galaxy are of the order of kiloyears or less for particles with rigidities of 10 EV \citep{Alvarez-Muniz:2001wyh,Murase:2008sa,Takami:2011nn}, while more recent GMF models produce significantly larger values dependent on the incident direction. On the other hand, many examples of multi-messenger sources exhibit variations on very short timescales \citep[e.g.][]{Funk:2015ena}, and bursting sources are believed to significantly contribute to the flux of UHECR \citep{AlvesBatista:2019tlv,Marafico:2024qgh,Farrar:2024zsm}.
Although it is widely accepted that potential UHECR sources must exhibit at least some degree of temporal variation, and while it has been shown that propagation in the extragalactic magnetic field has significant effects on observables, e.g.\ \cite{PierreAuger:2024hlp}, temporal effects of propagation in the galactic magnetic field have not yet been considered, to our knowledge. 

This work addresses this gap in our understanding of the GMF effects by extending simulations to rigidities below the ankle and accounting for the incidence positions and residence times in the Galaxy. This study is based on numerical simulations using CRPropa\,3.2 \cite{AlvesBatista:2016vpy,CRPropa:2022ovg}, and the robustness of our results is verified by employing multiple GMF models (JF12 \citep{Jansson:2012pc}, UF23 \citep{Unger:2023lob} and KST24 \citep{Korochkin:2024yit}).

We present a novel framework that connects the flux at the edge of the Galaxy to that observed on Earth. This framework is analogous to the widely used lensing technique but accounts for temporal effects and the initial incidence positions at the edge of the Galaxy. It can produce the observed spectrum, time dependence, arrival direction, etc., at Earth, from an arbitrary flux at the edge of the Galaxy. Like the lensing technique, it is an efficient method that uses precomputed simulation data obtained with Monte Carlo to model any desired scenario.

\section{The Galactic Response to Extragalactic Cosmic Ray Injection}
\subsection{Simulation Setup}
\label{sec:response}

The trajectories of extragalactic cosmic rays in the Galaxy depend not only on their rigidity but also on their entrance position and incoming direction. We characterize the galactic response to extragalactic cosmic rays by determining the distribution of propagation distances within the Galaxy until detection on Earth as a function of the rigidity, direction, and position at the edge of the Galaxy. We construct the distribution by sampling extragalactic cosmic rays from a distance sufficiently close to the Galaxy so that the propagation is only affected by the GMF. 

In CRPropa, we employ JF12, specifically the Planck-tuned version \cite{Jansson:2012pc,Planck:2016gdp}, together with a random realization of the corresponding turbulent field.
We inject extragalactic cosmic rays from a sphere of 20\,kpc radius centered on the Galactic center. They are propagated using the Cash-Karp algorithm implemented in CRPropa. For each particle, a random position is selected on the injection sphere and the initial direction into the Galaxy is sampled from a Lambertian distribution, using the CRPropa inbuilt function LambertDistributionOnSphere. In detail, the code aligns the local $z$ axis with the radius of the sphere (pointing to the center) and samples the initial direction from a probability distribution $P(\Omega) \sim \cos(\theta) d\Omega$ per solid angle $\Omega$ and angle $\theta$ measured from the local $z$ axis.
However, this distribution is also suitable for modeling the flux from an individual distant source.
The typical distances from sources are large enough that the flux from an individual source can be approximated as a plane wave. This can be achieved by selecting cosmic rays coming from the source direction. For a full derivation of this use case, see appendix \ref{app:lambert}, or e.g.\ \cite{lester:1979} for the general principle.

Unlike e.g.\ Ref.\,\cite{Farrar:2017lhm}, we use forward propagation in our simulations for several reasons. Computing the flux observed at Earth from the one at the edge of the Galaxy requires a prior distribution at the edge of the Galaxy with known properties that reflect physically motivated conditions, in our case either isotropic or distant point sources. Using back-propagation is ill suited for this as it may not be possible to ensure the desired distribution at the edge of the Galaxy nor its physical implications. On the other hand, the region of the phase space of initial parameters (positions, directions, rigidities) corresponding to non-observed events cannot be probed with back-propagation, as it samples only observed events, thus, the relative frequency of observed to non-observed events cannot be quantified. The back-propagation method is efficient because it ignores a large region of non-observed events. However, it comes at the price of assigning incorrect weights to observed events which arrived from realistic sources to the edge of the Galaxy. 

To ensure comparable convergence over multiple orders of magnitude, we simulated protons with a power-law rigidity distribution proportional to $R^{-1}$ from 10\,PV to 100\,EV. This range includes previously studied values and extends below the ankle. The cosmic rays were simulated until one of the following stopping conditions was met: a) arrival in the vicinity of the Earth (crossing the observer sphere), b) reaching the injection sphere, or c) propagating for a predefined maximum distance $L_{\rm max}=100$\,kpc.
The observer sphere is placed at (-8.5\,kpc, 0 ,0) in a cartesian coordinate system where the Galactic Center is at (0,0,0), the $x$-axis connects the Sun and the Galactic Center, and the $z$-axis points to the Galactic North pole. The radius is set to 1\,kpc.
Besides the reduced statistics, the size of the observer sphere has a negligible impact on the observed distribution \cite{Kaapa:2022tey}. 
The choice of the maximum propagation distance primarily affects our ability to neglect hadronic interactions of UHECR in the Galaxy. Assuming a gas density of one proton per cubic centimeter (which is a good approximation for the density in the Galactic plane \citep[see e.g.][]{Longair_2011}), the mean interaction length for protons is on the order of megaparsecs. For nuclei, it ranges from megaparsecs to sub-megaparsecs, depending on their mass. 

\begin{figure}[tbh]
    \centering 
    \includegraphics[width=\linewidth]{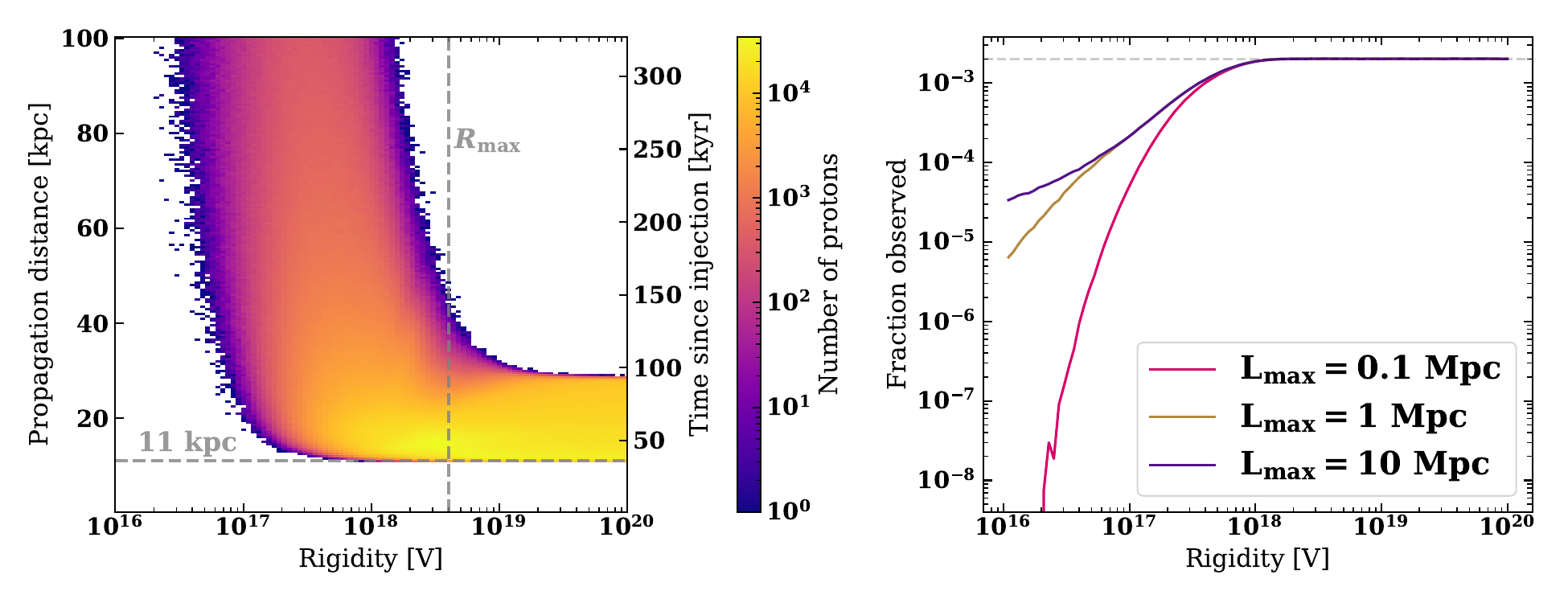}
    \caption{Left:
    Two‐dimensional event density histogram as a function of rigidity $R$ and propagation length $L$ in the Galaxy. The right axis expresses the corresponding propagation time, $t=L/c$. The dashed gray horizontal line marks the minimum distance between the injection and observer spheres, 11\,kpc. The dashed gray vertical line, $R_{\rm max}$, marks the typical value of maximal UHECR source rigidities, inferred from combined fits to the observed UHECR spectrum and composition \citep[e.g.][]{PierreAuger:2023htc}.
    Right: Relative acceptance, compared to the high-rigidity value, as a function of rigidity for different values of cutoff distance.}
    \label{fig:rig_vs_dist}
\end{figure}

\subsection{Characteristics of the Response Kernel}
The resulting distribution of galactic propagation lengths and times as a function of rigidity is shown in Fig.\,\ref{fig:rig_vs_dist} (left). The simulation involved 27 billion protons, of which approximately 32.7 million were recorded. This results in a total detection efficiency of $\frac{32.7\,10^{6}}{27\,10^{9}}=0.0012$, or 0.12~\%. As seen in Fig.\,\ref{fig:rig_vs_dist}, the losses due to $L_{\rm max}$ occur relatively more frequently at lower rigidities than at higher rigidities. Cosmic rays at the highest rigidities exhibit a time spread that matches geometric expectations from rectilinear propagation within the Galaxy. The Sun's off-center location explains the limits of 11~kpc for the closest point and 28~kpc for the farthest point from the injection sphere. In this rigidity limit, the acceptance is $\sim 0.2$~\% which is consistent with the expected geometrical acceptance assuming no GMF. 

The effects of the GMF gradually increase at $R \lesssim 10$~EV, first marginally extending the upper band of the propagation length distribution, until $R \lesssim 1$~EV. At this point, the spread in propagation lengths exceeds the limit distance $L_{\rm max} = 100$~kpc. Below this rigidity, the acceptance decreases (see right panel of Fig.~\ref{fig:rig_vs_dist}), partly because the fraction of the distribution with $L \geq L_{\rm max}$ is truncated. Previous simulations using $L_{\rm max} = 1$~Gpc showed that cosmic rays in the range $R=10^{16}-10^{17}$\,V spend a significant amount of time within the Galaxy \citep{Kaapa:2022qqn}, as can be seen in the right panel of Fig.\,\ref{fig:rig_vs_dist} for simulations with increased $L_{\rm max}$. However, the rapid decrease in acceptance towards lower rigidities cannot be explained solely by the truncation of the distributions. It is also affected by the size of the observer sphere. Smaller observer spheres generally require larger truncation values $L_{\rm max}$ to avoid artificially diluting the number of recorded particles at low rigidities \cite{Kaapa:2022tey}.  We note that the acceptance in the range $R=10^{17}-10^{18}$\,V decreases in the same way for $L_{\rm max} = 1$~Mpc as it does for $L_{\rm max} = 10$~Mpc. This indicates the stability of the simulation results, regardless of the  settings, and may indicate the influence of ``galactic shielding'' as discussed in Ref.\,\cite{Kaapa:2022qqn}.

The vertical gray line, $R_{\rm max}$, indicates the maximum rigidity value found in the combined fits of the UHECR spectrum and composition \cite[e.g.][]{PierreAuger:2016use,PierreAuger:2023htc}). 
Around this limit cosmic rays do not exhibit the ballistic behavior of the largest rigidities, which explains the increasingly complex behavior observed below 10~EV in Ref.\,\cite{Farrar:2017lhm}. 

The broad distribution of time since injection, or the age, of the observed cosmic rays (see right axis in Fig.\,\ref{fig:rig_vs_dist}, left) implies that, even in the scenario of a continuous and isotropic extragalactic flux, the observed flux contains a rigidity-dependent  mixture of cosmic rays of different ages, spanning from tens to hundreds of kiloyears. Note that the number or fraction of protons of a given age is contained in the horizontal sections of the 2D distribution. This has the immediate consequence that temporal changes in the extragalactic flux can produce features in the spectrum, which at the time of observation is composed of a mixture of injections from different epochs in the past. This possibility does not contradict Liouville's theorem, which only applies to stationary conditions, while temporal changes in flux intensity can produce features in the flux over limited time periods. 

\begin{figure}[t]
    \centering
        \includegraphics[width=0.49\linewidth]{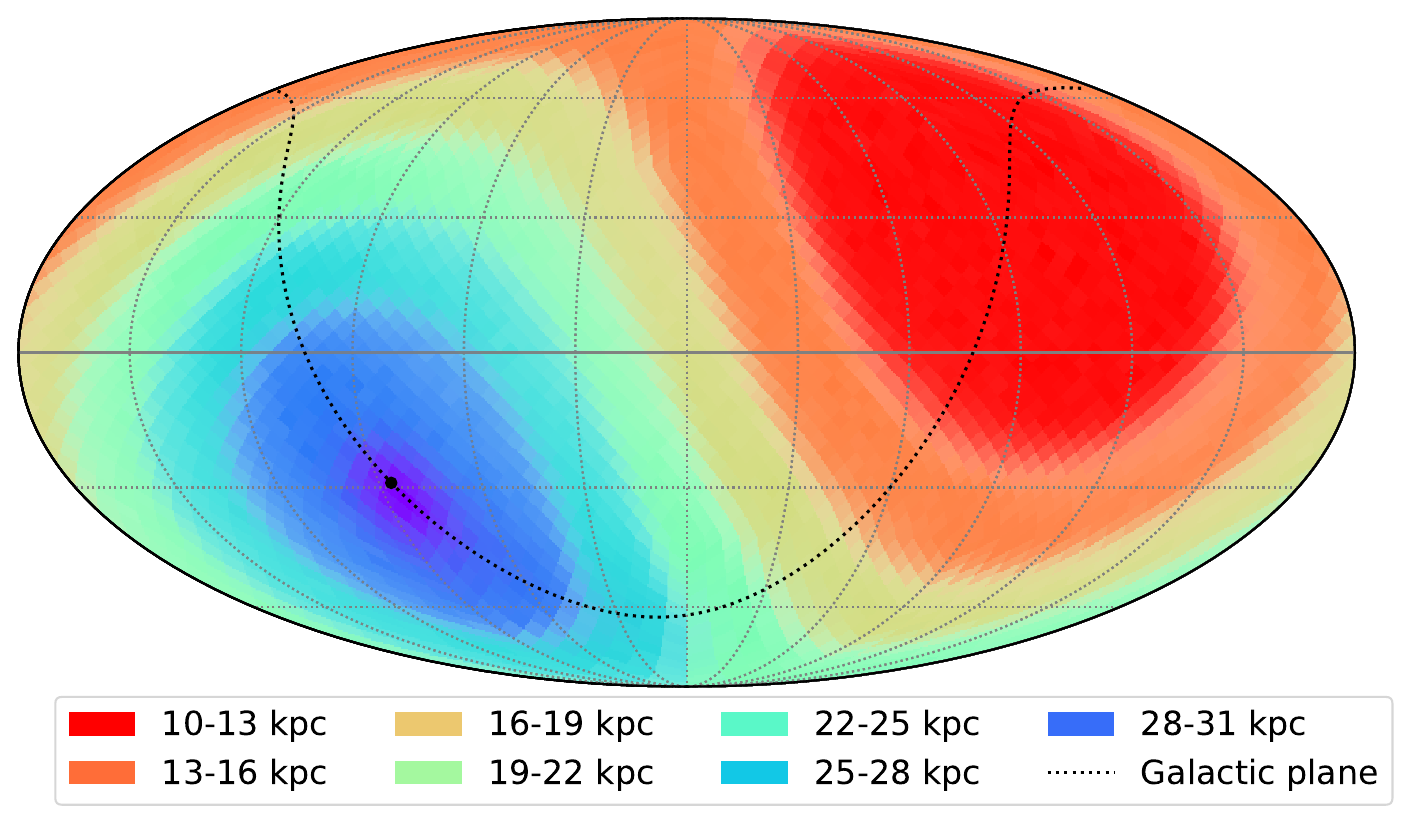} 
        \includegraphics[width=0.49\linewidth]{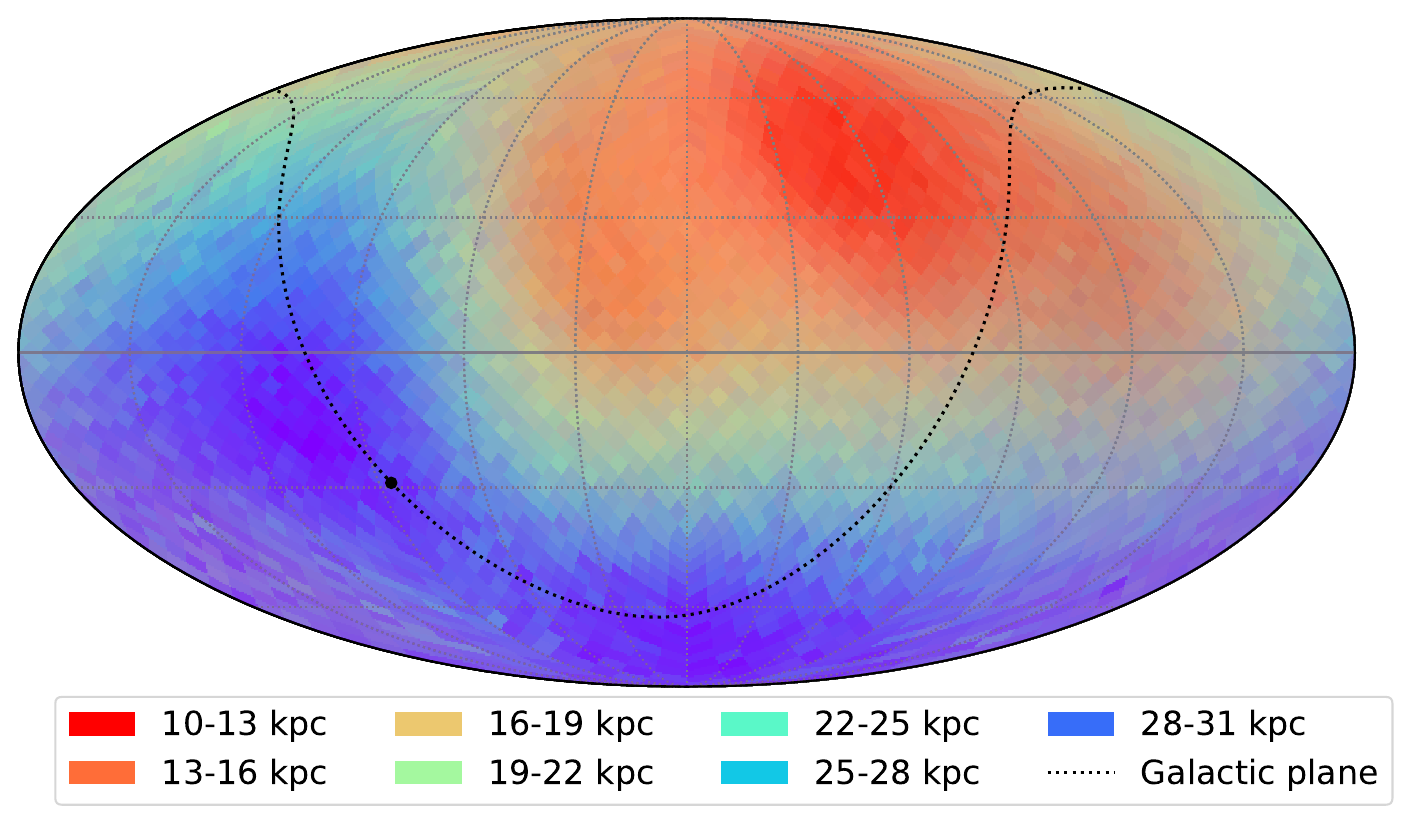}
    \caption{Distributions of cosmic ray propagation distances in the Galaxy as a function of the observed directions at Earth in equatorial coordinates. Left: $R>5$\,EV, Right: $R\in1-5$\,EV.}
    \label{fig:correlation_arrivaldirection_distance}
\end{figure}

UHECRs arriving from different directions in the sky have spent significantly different periods of time within the Galaxy before being observed.  
This is illustrated in Fig.\,\ref{fig:correlation_arrivaldirection_distance} (left) for cosmic rays with $R>5$\,EV. The color indicates the distance traveled in the Galaxy before observation. The clear structure confirms the geometric nature of the high-rigidity cosmic-ray sky: the largest distances correlate with the direction of the Galactic center, while the shortest distances correspond to the opposite direction. Intermediate distances correspond to concentric bands formed by specific angles relative to these main directions. Thus, the geometry of the simulation determines the results: arrival directions and age are highly correlated, and the minimum and maximal propagation distances are strictly bound. This geometric effect arises when the GMF's influence is negligible and thus disappears rapidly as rigidity decreases. Already in the 1-5\,EV range (see Fig.\,\ref{fig:correlation_arrivaldirection_distance} (right)), the arrival direction and the propagation distance are largely uncorrelated, especially for propagation distances in the Galaxy larger than the upper bound seen at high rigidities.

Other models of the GMF, besides JF12, are presented in Appendix\,\ref{app:otherB}. The variations caused by these models are very small and do not affect our conclusions.

\subsection{Temporal effects on the spectrum}
\label{sec:cutoff}

As seen in Fig.\,\ref{fig:rig_vs_dist} (left), the GMF quickly prolongs the trajectories of UHECRs with rigidities below 10 EV. This means that these particles remain in the Galaxy much longer than those with higher rigidities.

\begin{figure}[tbh]
    \centering
    \includegraphics[width=0.49\linewidth]{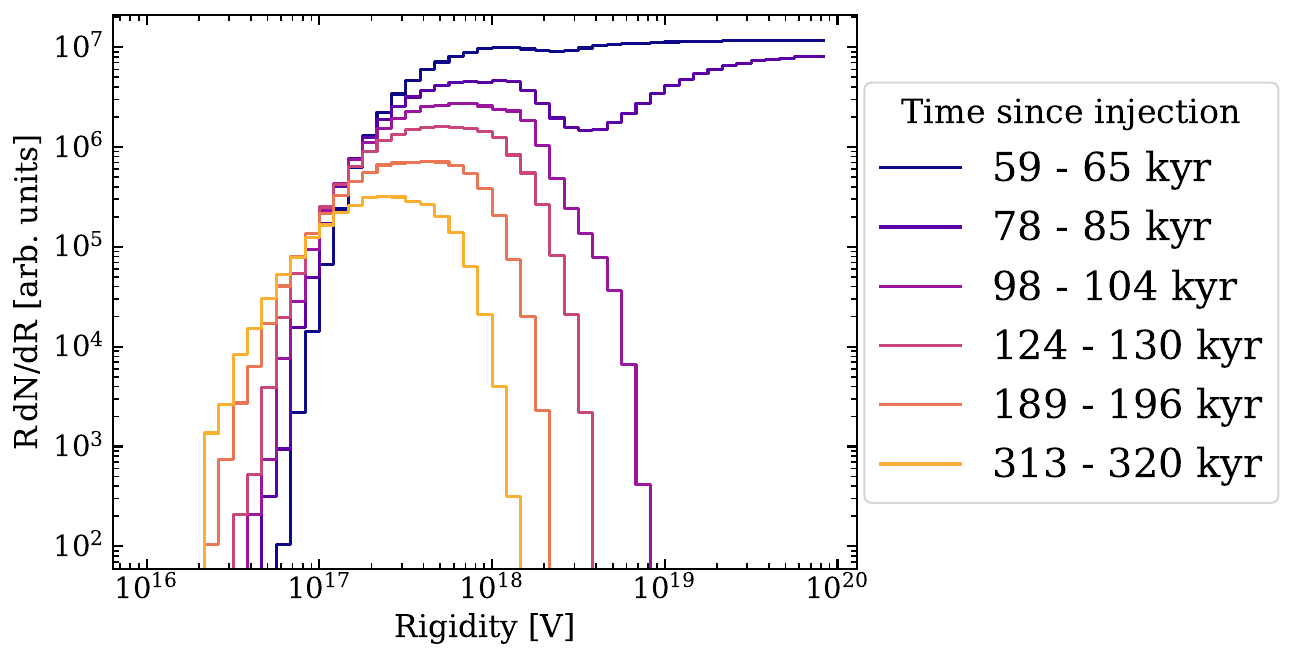}
    \includegraphics[width=0.49\linewidth]{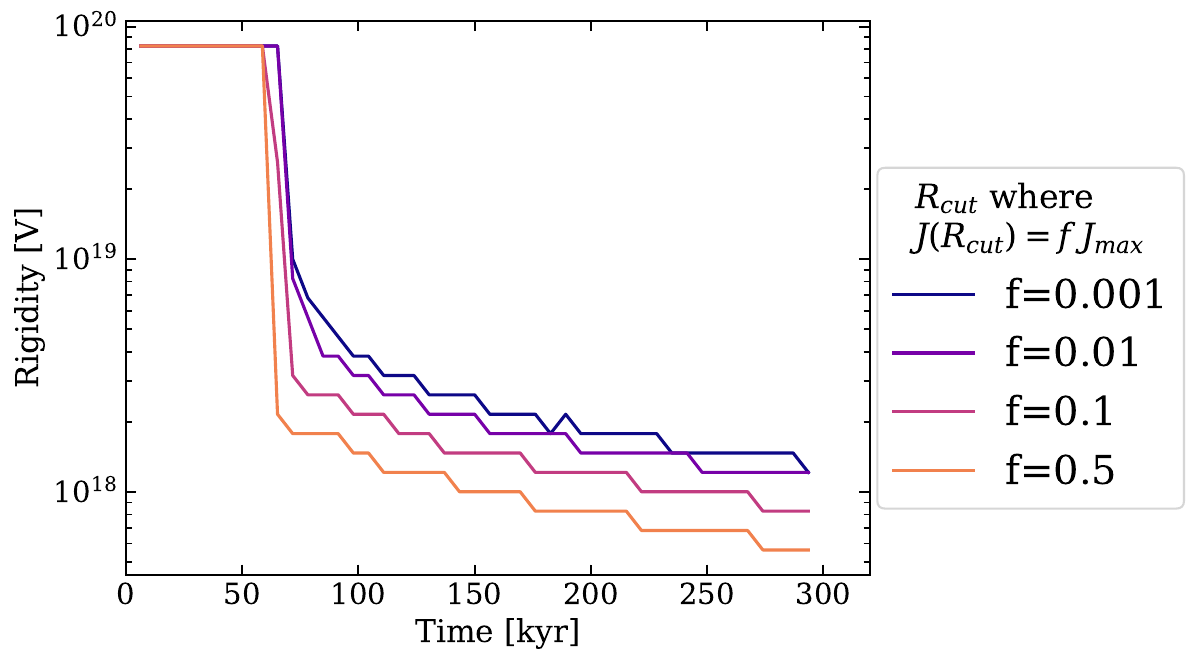}
       \caption[xx]{Left: Rigidity distribution of particles arriving at different times relative to the injection time at the edge of the Galaxy. 
       The spectra observed between about 100-300~kyr after an isotropic injection show a clear cutoff decreasing from approximately 5 to 1\,EV over time.
       Right: Temporal evolution of the cutoff. The plot shows the rigidity $R_{\rm cut}$ at which the flux $J$ drops to the percentage values of the maximum flux $J_{\rm max}$ denoted in the legend. }
       \label{fig:temporal_distribution}
\end{figure}

Now, consider an injection  infinitesimally narrow in time at the edge of the Galaxy from an isotropic distribution, following an $R^{-1}$ rigidity spectrum. Then, horizontal sections of the distribution in Fig.\,\ref{fig:rig_vs_dist} (left) represent the observed spectrum at the corresponding times since injection. Examples of these spectra are presented in Fig.\,\ref{fig:temporal_distribution} (left).

As can be seen, particles with high rigidities are present only for a short period after they have been injected, i.e.\ the very early spectra remain unchanged above 1 EV with respect to the injection. 
This matches the geometric expectation for particles traveling in straight or nearly straight lines.

The spectrum observed about 80\,kyrs after the injection shows an interesting dip between 1 and 10 EV. This depletion occurs because even particles with these rigidities, when entering from the opposite side of the Galaxy, travel through the bulk of the GMF experiencing some degree of deflections, which lengthens their trajectories. These particles are not lost but their longer paths make them arrive with later cohorts. The lower rigidity bump then consists of particles from all directions that were more significantly delayed.

Most importantly, all spectra observed at later times exhibit a strong cutoff, because the high-rigidity particles from the burst have already passed through the Galaxy. Lower-rigidity particles, however, remain trapped in the GMF and continue to be observed for a few hundred thousand years. This creates a time-varying cutoff in the spectra. The temporal dependence is shown in the right panel of Fig.\,\ref{fig:temporal_distribution}.

At any time $t \gtrsim 100$\,kyr, the cutoff appears below the GZK threshold \cite{Greisen:1966jv,Zatsepin:1966jv} for protons. Its rigidity values coincide with the maximum rigidity values of UHECR sources derived from combined fits to data from the Pierre Auger Observatory \cite{PierreAuger:2016use,PierreAuger:2023htc}. 
In the present scenario, however, the cutoff does not result from the UHECR sources, but rather from the temporal response of the Galaxy. It naturally occurs over extended periods after an initial burst, regardless of the maximal rigidity of UHECR sources. Since the cutoff is now a feature of the GMF, it also provides a natural explanation for the surprisingly narrow distribution of the maximum rigidity required to describe the UHECR spectrum and composition. In other words, the temporal effects of the GMF relax the constraint that different UHECR sources must have nearly the same maximum rigidity.
As can be seen in the left panel of Fig.~\ref{fig:temporal_distribution}, an original $R^{-1}$ spectrum unfolds into different components over time due to the temporal effects caused by the GMF. In agreement with Liouville's theorem, if the spectrum is integrated over all time, the modification disappears (see Fig.\,\ref{fig:rig_vs_dist}, right). 
However, temporal effects are not negligible and need to be considered when studying the observed UHECR spectrum.

In the 0.1 - 1 EV region, the spectra show a plateau that gradually decreases in intensity and shifts slowly to lower rigidity with increasing time since injection. 
In the range of 10 - 100 PV the lower bound of the temporal delays increase rapidly as well as the range of the distributions, reflecting the possible onset of diffusive propagation as the gyroradii fall below 100\,pc. 
The long temporal tails in Fig.~\ref{fig:rig_vs_dist} increase substantially the probability that these cosmic rays interact in the Galaxy decreasing the chances of being observed unaltered since injection. If extragalactic cosmic rays are significant at these rigidities, a proper treatment would require also to include their interactions to account for the flux of secondaries, which are not represented in the temporal distribution in Fig.~\ref{fig:rig_vs_dist}.

\section{Modeling the Galactic Response to Multiple Transient Injections}
\label{sec:MultipleInjections}

In this section, we approach a more realistic scenario: a supposed series of extragalactic bursts entering the Galaxy from different directions. We demonstrate how the galactic effects influence UHECR observables. A short burst results in an observed spectrum that evolves over time, differing from its original profile at each stage. A series of such bursts can produce a nearly continuous background of lower-rigidity particles, a cutoff for the highest rigidities, and various spectral features. The presence of these features depends on how much time has elapsed since the individual injections.

\subsection{A simplified burst scenario}
\label{sec:bursts}

Using a simplified toy model, we study how the recent history of extragalactic bursting sources could influence the currently observed features of UHECR due to modulation by the Galactic magnetic field. In this model, short-duration bursts of UHECRs are injected into the Galaxy at various times and from different directions.

Since the goal here is to demonstrate the concept rather than to tune the simulation results to observational data, the incoming directions were chosen randomly. Particular sources and realizations of the EGMF are subject of future work. 
%
For simplicity, we assume that all bursts inject the same number of UHECRs over a period of $\sim$70\,kyr, which is the suggested time span for UHECRs to arrive from nearby sources after propagating through the extragalactic magnetic field (EGMF) (see for example \citep{Marafico:2024qgh}). 

\begin{figure}[tbh]
    \centering
    \includegraphics[width=0.9\linewidth]{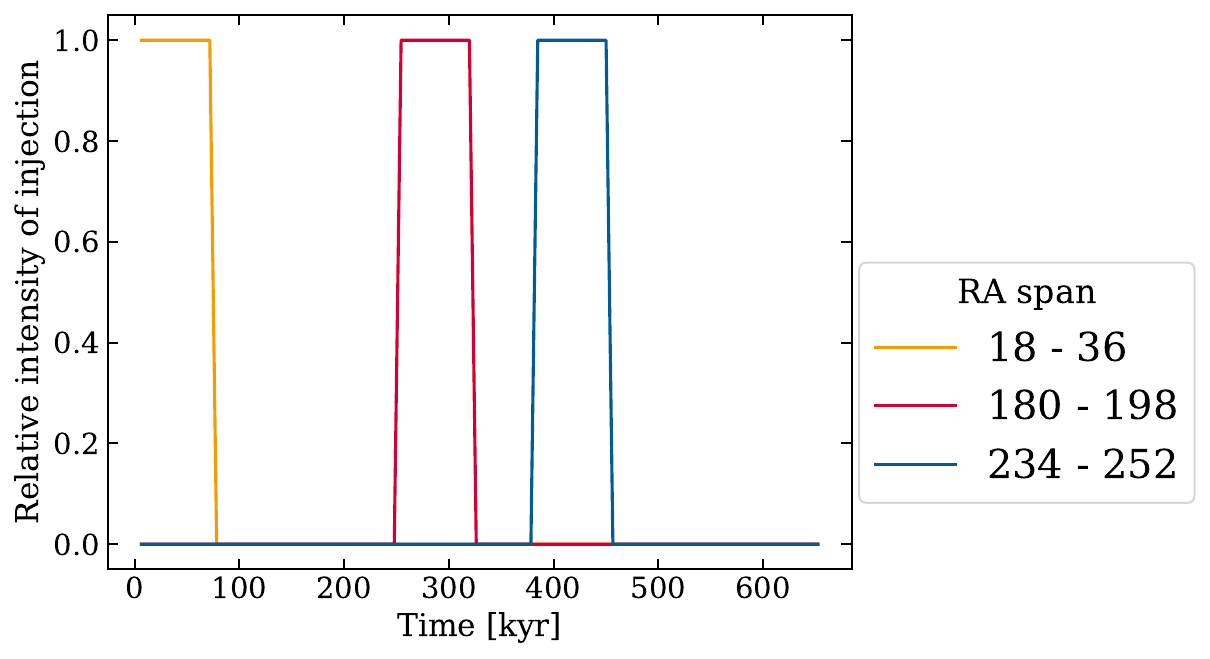}
    \caption{
    Example pulse profile for a pulse duration of  71.8\,kyr, showing three distinct burst events (vertical peaks), each color‐coded by the right ascension at which the source is located. The abscissa denotes the elapsed time since injection in kyr, and the ordinate shows the normalized luminosity, which is assumed equal for all bursting events for simplicity.
    }
    \label{fig:pulses}
\end{figure}

The average time interval between the UHECR burst injections can be estimated from the apparent density of sources and UHECR luminosity constraints, see e.g.\ \cite{AlvesBatista:2019tlv}. Based on the high level of observed UHECR isotropy at EeV energies, the Pierre Auger Collaboration derived source densities of $n_s \approx 10^{-4}$\,Mpc$^{-3}$ \cite{PierreAuger:2013waq}. The local burst rate can then be estimated by $\rho\approx n_s/\tau_d$, where $\tau_d$ is the apparent burst duration, which is not well known. Following this approach, the burst rate of e.g.\ HL-GRBs has been estimated at $\rho \approx$ (0.05-0.27)\,Gpc$^{-3}$\,yr$^{-1}$ \cite{Murase:2008sa,Takami:2011nn}, see also \cite{Marafico:2024qgh,AlvesBatista:2019tlv}. Within 30\,Mpc, a distance at which a source could significantly contribute to the local UHECR flux, we would then expect one burst every $\approx 100$\,kyr. A similar result was obtained in \citep{AlvesBatista:2019tlv}. This estimate has guided us to randomly place the three bursts in Fig.\,\ref{fig:pulses}. 

Our framework can easily model the response to single bursting sources. This is done by selecting events from the tensor in Fig.\,\,\ref{fig:rig_vs_dist}, that were injected from specific initial directions $\alpha_0$ (RA) and $\delta_0$ (declination) at any point $x_0$, $y_0$, and $z_0$ of the injection sphere. Thus, for an individual source, the arrival positions would cover the entire hemisphere of the starting sphere of our simulated cosmic rays while only the arrival direction at the galactic edge would be restricted to the source's direction. This is discussed in more detail in Appendix \ref{sec:directional-injection}.
In a more realistic scenario, the width of the arriving beam would depend on the distance of the source and the corresponding angular dispersion caused by the EGMF. This dispersion requires the inclusion of a rigidity-dependent injection into the Galaxy. In our analysis, however, we avoid this to make the description of Galactic effects more transparent before increasing the complexity of the injection process.
In this context, we note that more distant bursting or steady sources could still contribute to a constant UHECR background, which this analysis also ignores.

Figure\,\ref{fig:pulses} shows an example pulse profile, which was generated by first defining a uniform distance grid of 51 points between 0 and 100\,kpc and converting light‐travel distances to time using $c=2.998\times10^8\,$m/s.  The initial RA, $\alpha_0$, was partitioned into 20 bins, each $18^\circ$ wide. Three illustrative pulses were injected at RA of $18-36^\circ$, $180-198^\circ$, $234-252^\circ$. In each RA bin, a burst was imposed in a 100‐step time series at indices corresponding to burst onsets of 0, 238, and 363\,kyr. 

The temporal contribution of the $k$-th burst to the observed UHECR flux is the result of convolution between its time profile and the response kernel
\begin{equation}
f_k(ct, R, \alpha, \delta, \alpha^k_0) \;=\; \bigl(q_k * f\bigr)(t)
\;=\; \int_{0}^{\infty} q_k(\tau) \, f(c(t - \tau), R, \alpha, \delta, \alpha^k_0)\, \mathrm{d}\tau \: .
\end{equation}
For sufficiently short bursts, the injection function can be assumed to be $q_k(\tau)= a_0\delta(\tau - t_k)$, where $a_0$ is an arbitrary constant. The RA of the burst is indicated by $\alpha^k_0$.

\begin{figure}[tbh]
    \centering
    \includegraphics[width=0.9\linewidth]{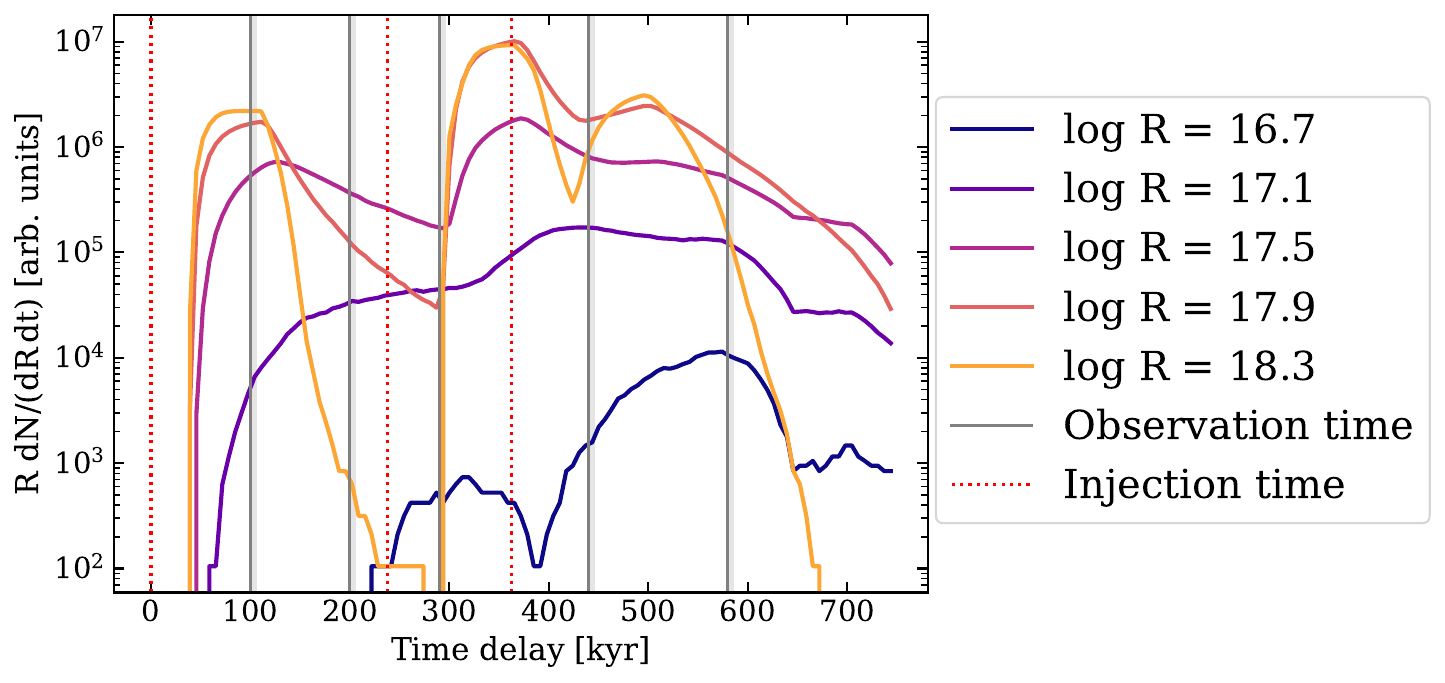}
    \caption{
    Time-dependent proton flux at Earth originating from the three short bursts of Fig.\,\ref{fig:pulses} injected at the edge of the Galaxy. Each colored curve represents the number of particles in a narrow rigidity bin, as labeled in the figure. The vertical red dotted lines mark the time of injection of each burst. The vertical gray lines indicate the five discrete observation windows used for sampling.
    }
    \label{fig:cr_time_evolution}
\end{figure}

Figure~\ref{fig:cr_time_evolution} displays the time evolution of the proton flux at the observer sphere, obtained by summing the contributions from all the bursts shown in Fig.\,\ref{fig:pulses}:
\begin{equation}
    J(t, R) = J_{0} \sum^3_{k=1} \int f_k(ct, R, \alpha, \delta, \alpha^k_0)\,  d\alpha \, d\delta \: .
\end{equation}

Each curve corresponds to a narrow rigidity bin, as indicated in the legend. The five gray vertical lines indicate $6\,\mathrm{kyr}$ wide windows that will be used for subsequent spectral analyses.
As expected from Fig.\,\ref{fig:rig_vs_dist}, the first peak (yellowish line) results from the nearly ballistic cosmic rays, which produce a sharp rise about 40~kyr after the first pulse injected at $t=0$ (red vertical dotted line). The lower the rigidity, the later the particles tend to arrive and the longer they tend to stay, i.e., they continue to arrive at the observer sphere. This demonstrates that the GMF effects can produce sustained cosmic‐ray fluxes at EeV energies and below from a few discrete injections over much larger timescales.

In this exploratory model, we assign the same intensity to all incoming bursts. In reality, however, different intensities are expected, particularly for sources located at different distances. This will inevitably influence the obtained spectra, since the profiles in Fig.\,\ref{fig:cr_time_evolution} are obtained by superposing fluxes from the various bursts. The unknown relative contributions of each burst increase the number of possible injection histories, any of which could result in the current observed state.

\subsection{UHECR Spectra}

\begin{figure}[tbh]
    \centering
    \includegraphics[width=\linewidth]{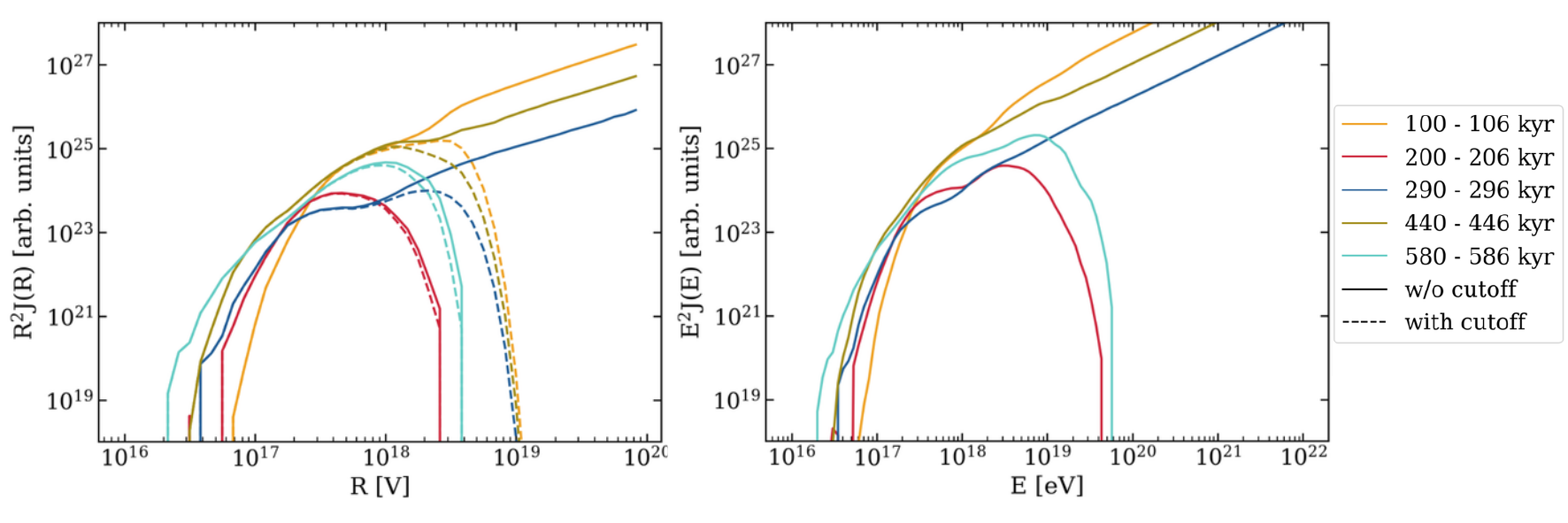}

    \caption{
    Left: Proton flux, $R^2 J(R)$, as a function of rigidity $R$ at the observer sphere, shown for five $6\,\mathrm{kyr}$ sampling windows following injection. The abscissa shows particle rigidity $R$ in V on a logarithmic scale.
    The solid lines reflect the observed spectra following an injection with a power-law distribution of $dN/dR \propto R^{-1}$, while the dashed lines apply a spectral cutoff at $2.5\,\mathrm{EV}$ (see Eq.\,\ref{eq:cutoff}). 
    Right: The same spectrum as a function of energy, assuming a mass composition as detailed in Table\,\ref{tab:injectedCompo}.
     }
    \label{fig:rigidity_spectra}
\end{figure}

Figure~\ref{fig:rigidity_spectra} shows the resulting cosmic ray flux at the observer sphere, sampled in the five contiguous $6\,$kyr observation time windows. 
Each curve corresponds to the average flux over the observation period
\begin{equation}
J(R)\;= \frac{1}{6\,{\rm kyr}}\int_{t_{\rm obs}}^{t_{\rm obs}+6\,{\rm kyr}}J(t, R)\,dt \: .
\end{equation}

The solid lines in the left panel of Fig.~\ref{fig:rigidity_spectra} represent the spectra taken at different observation times, as indicated in the legend. Spectra resulting from observations closely following an injection (the first, third, and fourth) yield very hard spectra up to the highest energies. However, spectra resulting from bursts observed $\sim$100\,kyr or more after the original injection are softer and exhibit a cutoff. This is because the highest-rigidity particles stream through the Galaxy on nearly straight lines and escape shortly after the burst.  

The hard tails of the spectra do not align with the observations of TA and the Auger Observatory. However, in our simulated scenario, they only appear shortly after a burst is injected into the Galaxy. To describe the observed UHECR spectra in combined fits to the energy spectrum and mass composition, an exponential or similar cutoff has been introduced to mark the maximum rigidity of the UHECR sources (see, e.g.\ \citep{PierreAuger:2016use}).
For comparison, the dashed lines show our results when applying a typical exponential cutoff of 
\begin{equation}
\label{eq:cutoff}
\frac{dN}{dR} \propto R^{-1}\cdot\exp\Bigl[-\Bigl(\frac{R}{R_{\rm max}}\Bigl)^2\Bigr] 
\end{equation}
to the injected spectrum, once $R>R_{\rm max}=2.5$\,EV. While such a cutoff affects the initial hard spectra, the later ones are largely unaffected, showing that the GMF effects give rise to the cutoff naturally. 
 
In this work, the maximum rigidity depends on the GMF and the time difference between when an extragalactic burst is injected into the Galaxy and when it is observed. Because high-rigidity cosmic rays can only be observed within a narrow time window, it is statistically much more probable to observe a spectrum that is cut off at rigidities of several EV.

The dip observed in the third time window (290 - 296\,kyr) has a slightly different origin than the dip described in Sec.\,\ref{sec:cutoff}. Here, it appears as a superposition of the (older) first burst and the earliest particles arriving from the (recent) second burst. While the origin of the spectral feature in Sec.\,\ref{sec:cutoff} was purely geometrical, a physically plausible scenario emerges here, where nuanced spectral features stem purely from temporal effects in the GMF.

\subsection{UHECR Composition}
\label{sec:composition}
Data from the Pierre Auger Observatory show that the composition of UHECR becomes increasingly heavy at energies above the ankle \cite{PierreAuger:2014gko,PierreAuger:2021fkf}. To examine how a superposition of particles with different rigidities at a given UHECR energy affects the conclusions of the present studies, we employ the kernel of Fig.\,\ref{fig:rig_vs_dist} and use the composition data as detailed in Table \ref{tab:injectedCompo}.

\begin{table}[tbh]
\centering
\begin{tabular}{|c|ccccc|}
\hline
Group &H&He&CNO&Si&Fe\\
\hline
Fraction & 0.63 & 0.37 & 0.06 & 0.003 & 0.0 \\
\hline
\end{tabular}
\caption{The composition that we injected is identical to the mass fractions at the sources published in \citep{PierreAuger:2016use}.
}
\label{tab:injectedCompo}
\end{table}

We approximate the energy of each UHECR mass group by $E_{\rm mass}=Z\times R$ and calculate the corresponding flux as
\begin{equation} 
J_{\rm mass}=J_{\rm orig} \times f_A \: .
\end{equation}
The total UHECR flux is obtained by summing up the fluxes of the different mass groups. The result is shown in the right panel of Fig.\,\ref{fig:rigidity_spectra} for illustration. In this simple study, the basic features of the spectra remain the same, with the cutoff in the all-particle spectrum appearing at a few $10^{19}$\,eV, again similar to observations \cite{PierreAuger:2020kuy,PierreAuger:2025eun}. This allows us to conclude that the energy-dependent composition does not influence our conclusions.

When a burst of extragalactic cosmic rays with a given composition is injected into the Galaxy, two effects will lead to a time variability in the composition observed at Earth over time. Note that the composition of UHECRs is assessed as a function of energy, $E$. According to $R=E/Z$ (where $Z$ is the atomic number), each bin of energy will consist of high-rigidity protons and low-rigidity nuclei. The first effect causing a time variability in the composition is the delayed arrival of the heavier nuclei due to their lower rigidity. This effect dominates during the onset of the burst. The second effect is related to the rigidity-dependent confinement time in the Galaxy. This causes high-rigidity protons to escape more rapidly than low-rigidity nuclei with the same energy. Both effects are visible in Fig.\,\ref{fig:cr_time_evolution}: Low-rigidity particles, i.e.\ nuclei, arrive later and stay longer in the Galaxy. Consequently, the composition at high energies becomes increasingly heavy over time since its injection into the Galaxy.
Table\,\ref{tab:composition} demonstrates this effect quantitatively at an energy of 1\,EeV. For clarity, we exclude the second and third burst in this discussion and only consider the evolution of the composition from a single burst arriving at the edge of the Galaxy at $t=0$ at RAs ranging from $18^\circ$ to $36^\circ$. The observation times are selected at 100, 200, and 300 kiloyears after its injection. Clearly, the composition becomes heavier over time. We remind the reader again that interactions are not included.

\begin{table}[tbh]
\centering
\begin{tabular}{|c|cccc|}
\hline
  &H&He&CNO&Si\\
\hline
100\,kyr & 0.68 & 0.28 & 0.04 & 0.0 \\
200\,kyr & 0.10 & 0.22 & 0.67 & 0.01 \\
300\,kyr & 0.02 & 0.09 & 0.84 & 0.06 \\
\hline
\end{tabular}
\caption{Composition fractions at 1 EeV for a single burst from RA $18-36^\circ$. The composition becomes heavier over time due to propagation in the GMF, because of the delays between rigidities and different confinement times.}
\label{tab:composition}
\end{table}

\subsection{Skymaps and Dipoles}
The anisotropy of UHECR is another important observable that is often used to learn about their origin, hypothesizing correlations between arrival directions and the positions of potential sources. To study the skymaps expected from our simple toy model scenario,  we compute the particle numbers as a function of right ascension, $\alpha$, and declination, $\delta$ by
\begin{equation}
S_{\rm obs}(\alpha, \delta)\;= \frac{J_{0}}{6\,{\rm kyr}}\int_{t_{\rm obs}}^{t_{\rm obs}+6\,{\rm kyr}}dt \sum^3_{k=1} \int f_k(ct, R, \alpha, \delta, \alpha^k_0) \, dR \: .
\end{equation}
To estimate the dipole amplitude, 
we rebin the skymaps into 12 right ascension bins and divide the values by the mean. We then fit the data points to a cosine function.

\begin{figure}[tbh]
    \centering
    \includegraphics[width=0.45\linewidth]{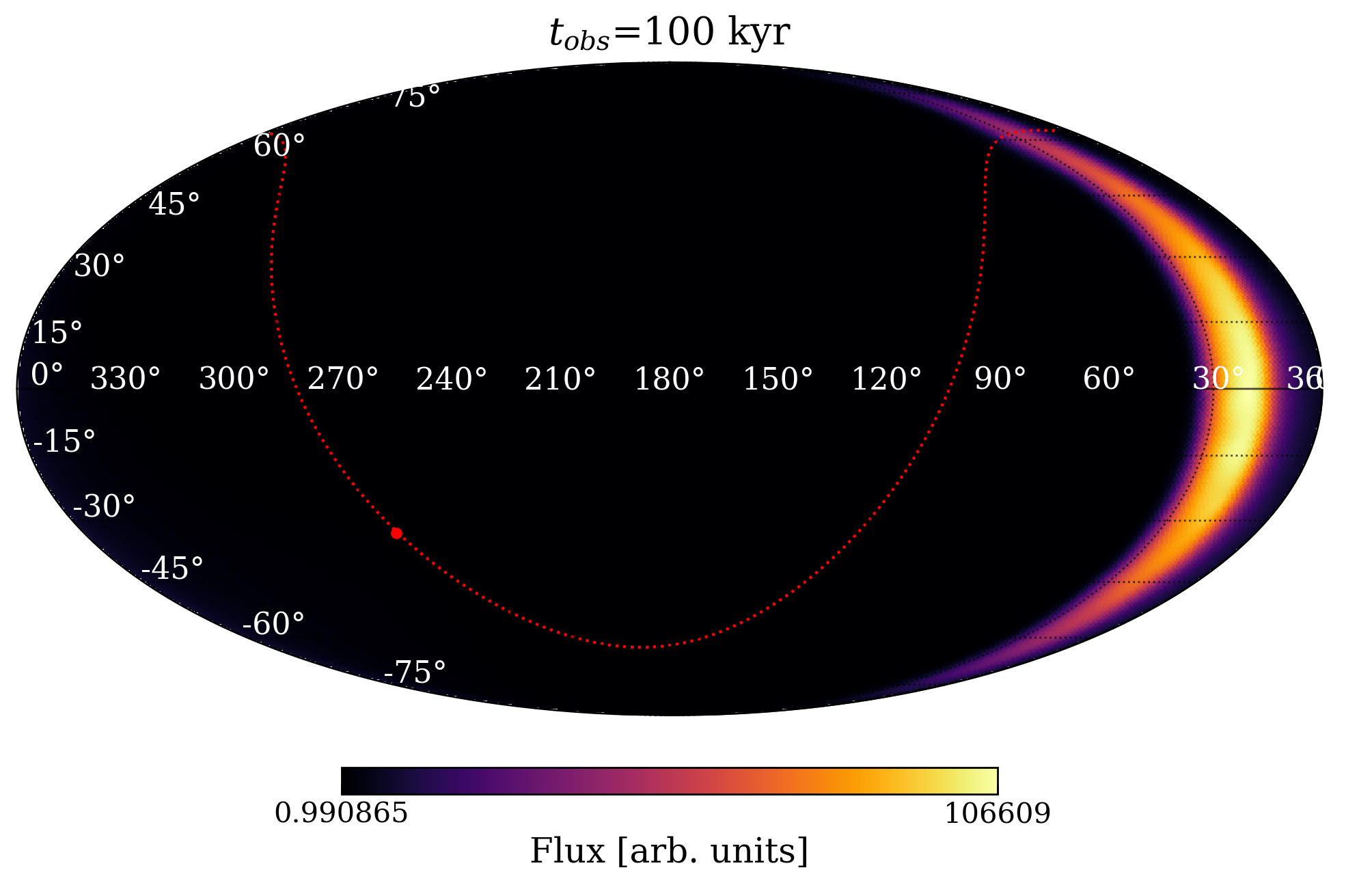}
        \includegraphics[width=0.4\linewidth]{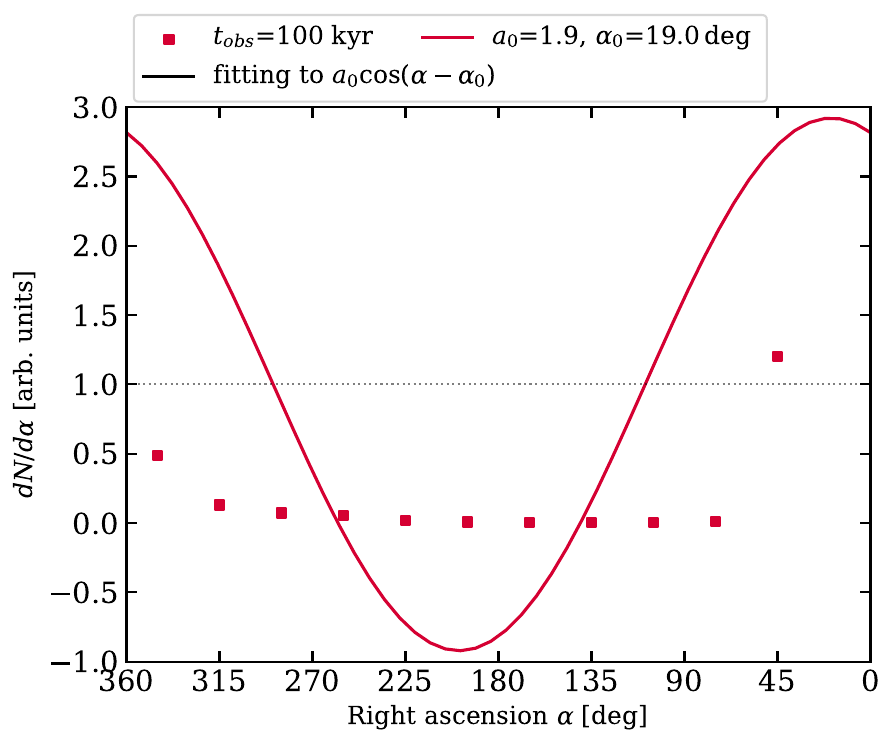}
    \caption{
    Left: Healpy Mollweide projection of the all‐sky arrival‐direction distribution of simulated cosmic‐ray protons at the observer sphere. The data cover the first observation window at 100 kyr and are shown in equatorial coordinates after integrating over all rigidities, assuming an injection spectrum into the Galaxy of $dN/dR/\propto R^{-1}$.  
    The galactic plane is indicated by the red dashed line.
    Right: Corresponding right‐ascension profile obtained by summing the same sky map in 12 equal RA bins and normalizing to a unit mean.  
    The points show the simulated counts per bin. The solid curve is a cosine fit with an amplitude $a_{0}=1.9$ and a phase of $\alpha_{0}=19^\circ$.
    }
    \label{fig:skymap}
    \centering
    \includegraphics[width=0.45\linewidth]{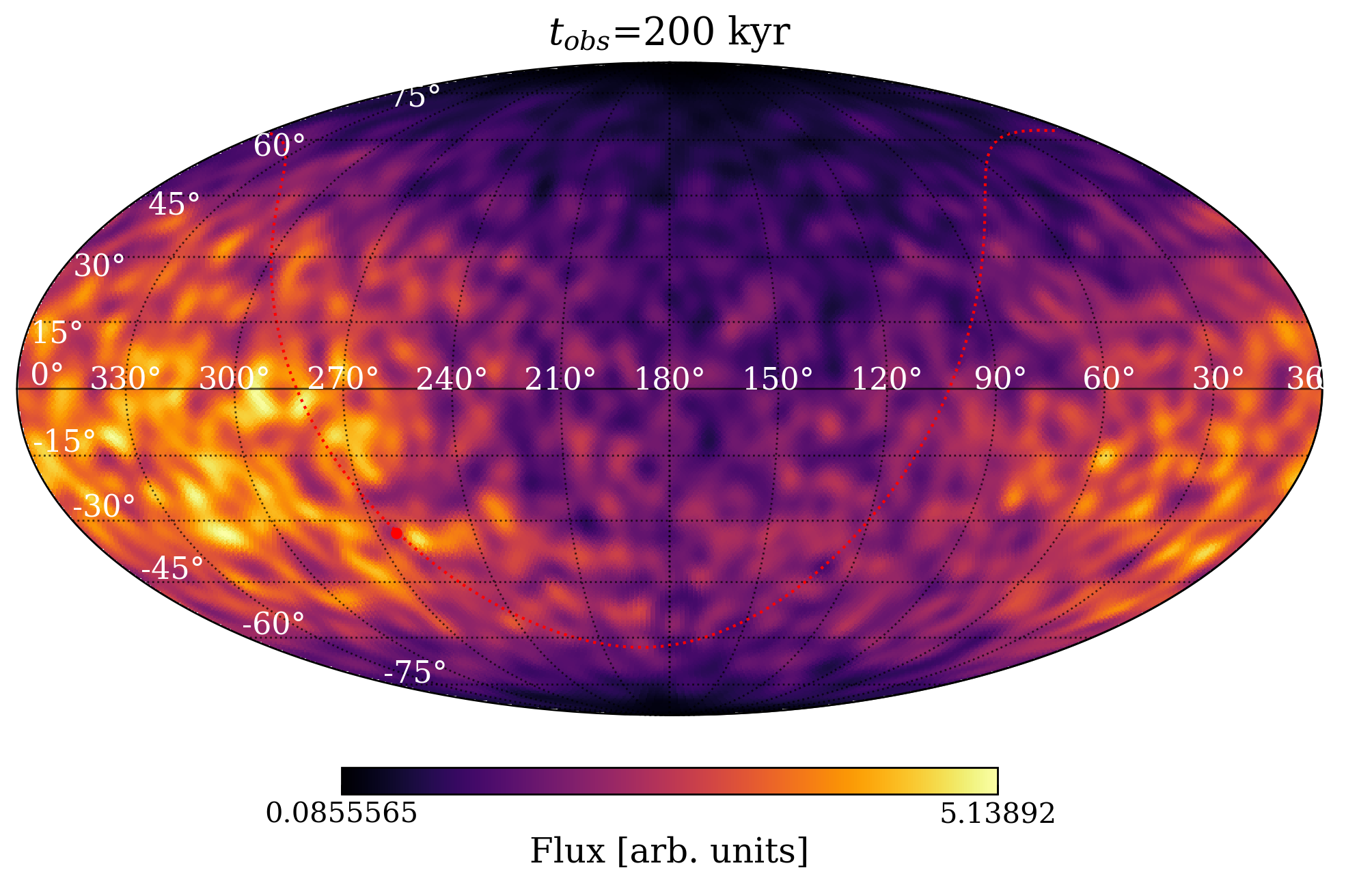}
    \includegraphics[width=0.4\linewidth]{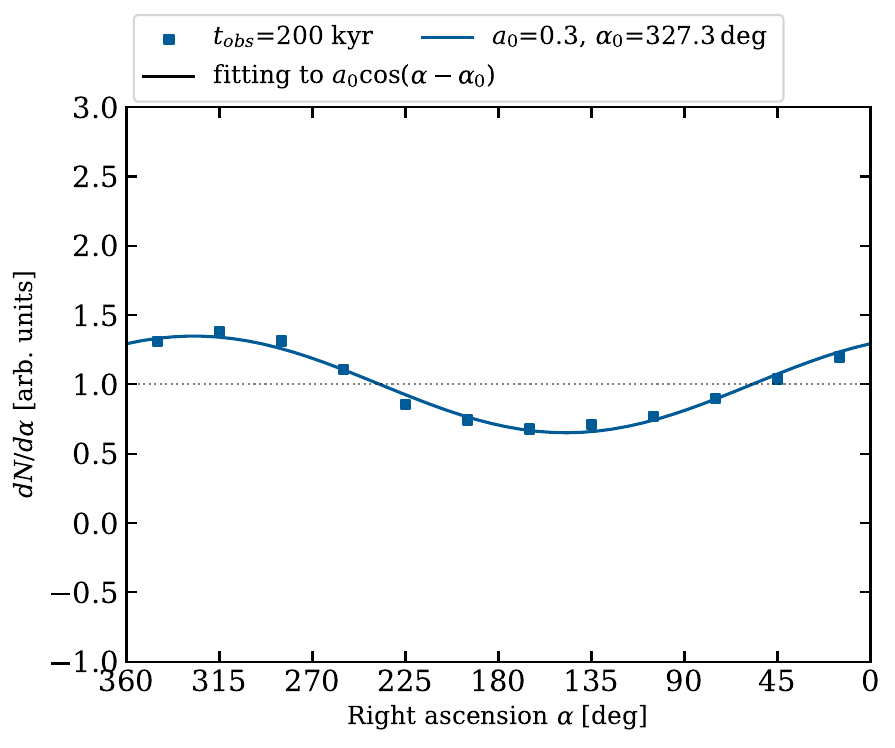}
    \caption{ 
    Same as Fig.\,\ref{fig:skymap}, but observed 100~kyr later, at 200~kyr. The data provide a very good fit to a dipole with an amplitude of $a_{0}=0.3$ and a phase shift of $\alpha_{0}=327.3^\circ$. A multipole expansion of the skymap also exhibits a quadrupole component, which is suppressed relative to the dipole by a factor of 3.1.
    }
    \label{fig:skymap_later}
\end{figure}

Shortly after injection, the resulting skymap is strongly dominated by the right ascension band of the injection, whereas the rest of the sky is dominated by delayed contributions from the earlier portion of the injection. 
This is because the injection spectrum chosen follows a $R^{-1}$ distribution and the high-rigidity particles travel in a almost straight lines, while, for lower rigidities, particles are dispersed over longer periods of time decreasing their intensity.
Summing up over all rigidities without further reweighting, the contributions per right ascension bin show a peak that does not fit well with a cosine, as shown in  Fig.\,\ref{fig:skymap}.

However, at later observation times, particles arrive from a much broader range of directions because the deflections in the GMF increasingly wash out the original direction. Figure\,\ref{fig:skymap_later} on the left shows that the particles appear to come from a broad region near the Galactic center. This contrasts with their original injection from a different direction outside the Galaxy. Furthermore, the resulting skymaps fit  very well to a dipole (\textit{c.f.}\ Fig.\,\ref{fig:skymap_later}, right). We note that the amplitude is larger than what has been measured by the Pierre Auger Observatory \citep{PierreAuger:2017pzq,PierreAuger:2024fgl}, which is  consistent with expectations for several reasons. First, the kernel in Fig.\,\ref{fig:rig_vs_dist} assumes a flat distribution in $\log(R)$ to allow reweighting when computing the observed flux for arbitrary injection spectra. The corresponding $R^{-1}$ spectral shape is much harder than the observed CR energy spectrum. Thus, the skymaps presented here are heavily influenced by the high-rigidity particles, which are known to exhibit larger dipole amplitudes \cite{PierreAuger:2024fgl}.
Second, in a more realistic scenario, we expect an isotropic UHECR background from distant sources. Third, we expect multiple injections into the Galaxy. All of these effects would reduce the simulated dipole amplitude. Moreover, isotropization increases over time relative to the injection period, which is a typical outcome of our simulations. 
Future studies aiming at a quantitative comparison to observational data will address these points by reweighting the rigidity spectra, converting them to energy while accounting for the UHECR mass composition, applying proper energy thresholds, and examining the directions and amplitudes of the dipoles arising from different injection scenarios.

At this point, we conclude that this model can produce dipole-like anisotropies. These appear to be a general feature of all the burst realizations that have been studied so far. However, many different injection histories could result in a similar outcome to the one discussed above. Therefore, the true time evolution of the UHECR flux into our Galaxy may not be inferable from the observed flux on Earth alone.

\subsection{Effects of the Extragalactic Magnetic Field}
\label{sec:egmf}
We adopt the assumptions regarding the effect of the EGMF on the temporal spread of UHECRs arriving from bursting sources, discussed in, for example \citep{Marafico:2024qgh}. We note that particles of all rigidities, especially the lower ones, do not necessarily reach the edge of the Galaxy simultaneously in a delta-like burst. Instead, they arrive delayed (see for example \citep{Miralda-Escude:1996twc,Mbarek:2025xvg}), similar to the effects within the GMF discussed in this work. Since the initial exploration presented here is intended to illustrate the galactic effect, we do not explicitly account for this spread. However, since we are interested in nearby sources within a distance of about 30\,Mpc or less, this spread is not expected to exceed the assumed time spread of 70\,kyr \cite{Stanev:2000fb}.

Furthermore, as shown in \citep{Eichmann:2022ias}, for individual bursting sources to contribute significantly at 10~EV, the escape time should be of order $\sim 10^5$~kyr. Otherwise, the source's magnetic fields can smear the emission over a longer period below a certain rigidity.


\section{Conclusions}

In this study, we demonstrate that the GMF introduces arrival time delays that depend on the rigidity of the particles. 
These delays alter the observed properties of extragalactic cosmic rays with respect to their injection at the edge of the Galaxy. Most importantly, they alter the energy spectrum and cosmic ray anisotropies in a time-dependent way. 
The findings of this work are robust against the choice of the GMF model and are mostly dependent on the overall strength of the GMF and the geometry (\textit{c.f.} Appendix \ref{app:otherB}). The main points can be summarized as follows:

\begin{itemize}
    \item A significant feature that emerges is a cutoff in the energy spectrum of extragalactic cosmic rays, which appears in the rigidity range between $10^{18}-10^{19}$~V.
    \item This cutoff is observed over an extended period of time after a burst of extragalactic cosmic rays has reached the edge of our Galaxy and is at a level where distinct injections from sources can be resolved. 
    \item The mass composition is modified, compared to the edge of the Galaxy, with heavier elements becoming more dominant over time. 
    \item A dipole-like anisotropy can arise from such delays on timescales of a hundred kilo-years.
\end{itemize}

This paradigm offers a completely different perspective on UHECR observations. Most importantly, it provides an alternative explanation for the origin of the cutoff observed in the UHECR spectrum, eliminating the need for an acceleration limit in extragalactic sources in bursting scenarios. This limit was referred to as ``the disappointing model'' in \cite{Aloisio:2009sj}. It could also resolve the ``curious case of the maximum rigidity distribution of cosmic-ray accelerators'', pointed out in \cite{Ehlert:2022jmy} where it was found that the maximum rigidities of extragalactic sources must be identical within a factor of three. 

Interpreting the cutoff as a feature of the GMF could also alter constraints derived from this assumption. 

For example, it creates a new opportunity for the future detection of high-energy cosmogenic neutrinos. The maximum source rigidity derived from the combined fits to the UHECR energy spectra and composition leads to cosmogenic neutrino fluxes that are unlikely to be detected, even by the next generation of neutrino observatories \cite{PierreAuger:2019ens,AlvesBatista:2019tlv,Wittkowski:2018giy,Coleman:2022abf}. The scenario discussed here, would allow the maximum source rigidity to be much higher than a few EV. This would enable efficient photo-pion and photodisintegration interactions of UHECR with the CMB, thereby raising the cosmogenic neutrino fluxes considerably. However, a quantitative calculation must account for the burst frequency and cosmological evolution of transient UHECR and neutrino sources. Similar to UHECR, UHE neutrinos from transient sources will quickly pass through the Galaxy. However, unlike UHECRs, neutrinos will reach us from cosmological distances. This results in a steadier flux from many transient events. 

Finally, we stress again that due to escape times from the sources and propagation in the EGMF, it is unlikely that particles of all rigidities arrive to the Milky Way at the same time. We adopted this simplified approach regardless for two reasons: one is consistency, as this assumption is prevalent in the existing literature. The other is that our focus is on describing the intrinsic effects caused by the GMF, which will persist under any injection, but are more clearly understood in the most streamlined scenarios. 

We also reiterate that this study is exploratory and does not aim to quantitatively describe all UHECR observational features. The chosen parameters, such as the burst frequency and duration, are well-motivated by the literature, while the injection directions were chosen arbitrarily to illustrate their effects. Future work will address the features and temporal changes of the observed cosmic ray mass composition and their anisotropies more quantitatively and will also account for rigidity dependent time-lags of extragalactic cosmic rays arriving at the edge of the Galaxy with the aim to constrain specific scenarios of UHECR sources.

\paragraph{Data availability:} The authors will make the data and results of this study available upon email request.

\section*{Acknowledgments}
This work was supported by the Deutsche Forschungsgemeinschaft (DFG) through the Collaborative Research Center SFB 1491 "Cosmic Interacting Matters - From Source to Signal" (project no.\ 445052434) as well as the "MultI-messenger probe of Cosmic Ray Origins" (MICRO) grant (project no.\ 445990517, KA 710), and by the BMFTR Verbundforschung Astroteilchenphysik. The computations were partially carried out on the PLEIADES cluster at the University of Wuppertal, supported by the DFG (grant no.~INST 218/78-1 FUGG).

\bibliographystyle{unsrtnat}
\bibliography{references}

@misc{crpropadocs:lambert,
    key = "CRPropa 3.2 Documentation", 
    url = {https://crpropa.github.io/CRPropa3/api/classcrpropa_1_1Random.html#_CPPv4N7crpropa6Random18randVectorLambertsEv} ,
     note = {Accessed: 2026-08-20}
}

@article{PierreAuger:2021fkf,
    author = "Aab, Alexander and others",
    collaboration = "Pierre Auger",
    title = "{Deep-learning based reconstruction of the shower maximum $X_{max}$ using the water-Cherenkov detectors of the Pierre Auger Observatory}",
    eprint = "2101.02946",
    archivePrefix = "arXiv",
    primaryClass = "astro-ph.IM",
    reportNumber = "FERMILAB-PUB-21-084-AD-AE-SCD-TD, FERMILAB-PUB-21-084-AD-AE-SCD-TD",
    doi = "10.1088/1748-0221/16/07/P07019",
    journal = "JINST",
    volume = "16",
    number = "07",
    pages = "P07019",
    year = "2021"
}

@article{lester:1979,
       author = {{Lester}, T.~P. and {McCall}, M.~L. and {Tatum}, J.~B.},
        title = "{Theory of Planetary Photometry}",
      journal = {JRASC},
         year = 1979,
        month = oct,
       volume = {73},
        pages = {233},
       adsurl = {https://ui.adsabs.harvard.edu/abs/1979JRASC..73..233L}
}

@article{Aloisio:2009sj,
    author = "Aloisio, R. and Berezinsky, V. and Gazizov, A.",
    title = "{Ultra High Energy Cosmic Rays: The disappointing model}",
    eprint = "0907.5194",
    archivePrefix = "arXiv",
    primaryClass = "astro-ph.HE",
    doi = "10.1016/j.astropartphys.2010.12.008",
    journal = "Astropart. Phys.",
    volume = "34",
    pages = "620--626",
    year = "2011"
}

@article{Zatsepin:1966jv,
    author = "Zatsepin, G. T. and Kuzmin, V. A.",
    title = "{Upper limit of the spectrum of cosmic rays}",
    journal = "JETP Lett.",
    volume = "4",
    pages = "78--80",
    year = "1966"
}

@article{Thielheim:1968,
    author = "Thielheim, K. O. and Langhoff, W.",
    title = "{Trajectories of high-energy cosmic rays in the galactic disk}",
    journal = "J. Phys. A: Gen Phys.",
    volume = "1",
    pages = "694--703",
    year = "1968",
    doi = {10.1088/0305-4470/1/6/308}
}

@article{Funk:2015ena,
    author = "Funk, Stefan",
    title = "{Ground- and Space-Based Gamma-Ray Astronomy}",
    eprint = "1508.05190",
    archivePrefix = "arXiv",
    primaryClass = "astro-ph.HE",
    doi = "10.1146/annurev-nucl-102014-022036",
    journal = "Ann. Rev. Nucl. Part. Sci.",
    volume = "65",
    pages = "245--277",
    year = "2015"
}

@article{PierreAuger:2016use,
    author = "Aab, Alexander and others",
    collaboration = "Pierre Auger",
    title = "{Combined fit of spectrum and composition data as measured by the Pierre Auger Observatory}",
    eprint = "1612.07155",
    archivePrefix = "arXiv",
    primaryClass = "astro-ph.HE",
    reportNumber = "FERMILAB-PUB-16-618",
    doi = "10.1088/1475-7516/2017/04/038",
    journal = "JCAP",
    volume = "04",
    pages = "038",
    year = "2017",
    note = "[Erratum: JCAP 03, E02 (2018)]"
}

@article{PierreAuger:2014gko,
    author = "Aab, A. and others",
    collaboration = "Pierre Auger",
    title = "{Depth of maximum of air-shower profiles at the Pierre Auger Observatory. II. Composition implications}",
    eprint = "1409.5083",
    archivePrefix = "arXiv",
    primaryClass = "astro-ph.HE",
    reportNumber = "FERMILAB-PUB-14-347-AD-AE-CD-TD",
    doi = "10.1103/PhysRevD.90.122006",
    journal = "Phys. Rev. D",
    volume = "90",
    number = "12",
    pages = "122006",
    year = "2014"
}

@article{Miralda-Escude:1996twc,
    author = "Miralda-Escude, Jordi and Waxman, Eli",
    title = "{Signatures of the origin of high-energy cosmic rays in cosmological gamma-ray bursts}",
    eprint = "astro-ph/9601012",
    archivePrefix = "arXiv",
    reportNumber = "IASSNS-AST-96-2",
    doi = "10.1086/310042",
    journal = "Astrophys. J. Lett.",
    volume = "462",
    pages = "L59--L62",
    year = "1996"
}

@article{Mbarek:2025xvg,
    author = "Mbarek, Rostom and Caprioli, Damiano",
    title = "{Revisiting propagation delays of ultra-high-energy cosmic rays from long-lived sources}",
    eprint = "2502.01022",
    archivePrefix = "arXiv",
    primaryClass = "astro-ph.HE",
    doi = "10.1016/j.astropartphys.2025.103148",
    journal = "Astropart. Phys.",
    volume = "172",
    pages = "103148",
    year = "2025"
}

@article{Korochkin:2024yit,
    author = "Korochkin, Alexander and Semikoz, Dmitri and Tinyakov, Peter",
    title = "{The coherent magnetic field of the Milky Way halo, the Local Bubble, and the Fan region}",
    eprint = "2407.02148",
    archivePrefix = "arXiv",
    primaryClass = "astro-ph.GA",
    doi = "10.1051/0004-6361/202451440",
    journal = "Astron. Astrophys.",
    volume = "693",
    pages = "A284",
    year = "2025"
}

@article{Wittkowski:2018giy,
    author = "Wittkowski, David and Kampert, Karl-Heinz",
    title = "{On the flux of high-energy cosmogenic neutrinos and the influence of the extragalactic magnetic field}",
    eprint = "1810.03769",
    archivePrefix = "arXiv",
    primaryClass = "astro-ph.HE",
    doi = "10.1093/mnrasl/slz083",
    journal = "Mon. Not. Roy. Astron. Soc.",
    volume = "488",
    number = "1",
    pages = "L119--L122",
    year = "2019"
}

@inproceedings{Alvarez-Muniz:2001wyh,
    author = "Alvarez-Muniz, J. and Engel, R. and Stanev, T.",
    title = "{Propagation of ultra-high energy cosmic rays in the galaxy}",
    booktitle = "{27th International Cosmic Ray Conference}",
    month = "8",
    year = "2001"
}

@article{Harari:1999it,
    author = "Harari, Diego and Mollerach, Silvia and Roulet, Esteban",
    title = "{The Toes of the ultrahigh-energy cosmic ray spectrum}",
    eprint = "astro-ph/9906309",
    archivePrefix = "arXiv",
    doi = "10.1088/1126-6708/1999/08/022",
    journal = "JHEP",
    volume = "08",
    pages = "022",
    year = "1999"
}

@book{Longair_2011, 
place={Cambridge}, 
edition={3}, 
title={High Energy Astrophyics}, 
publisher={Cambridge University Press}, 
author={Malcom S. Longair}, year={2011}}

@article{PierreAuger:2023htc,
    author = "Halim, A. Abdul and others",
    collaboration = "Pierre Auger",
    title = "{Constraining models for the origin of ultra-high-energy cosmic rays with a novel combined analysis of arrival directions, spectrum, and composition data measured at the Pierre Auger Observatory}",
    eprint = "2305.16693",
    archivePrefix = "arXiv",
    primaryClass = "astro-ph.HE",
    reportNumber = "FERMILAB-PUB-24-0135-AD-CSAID-PPD-TD-V",
    doi = "10.1088/1475-7516/2024/01/022",
    journal = "JCAP",
    volume = "01",
    pages = "022",
    year = "2024"
}

@article{Murase:2008sa,
    author = "Murase, Kohta and Takami, Hajime",
    title = "{Implications of Ultra-High-Energy Cosmic Rays for Transient Sources in the Auger Era}",
    eprint = "0810.1813",
    archivePrefix = "arXiv",
    primaryClass = "astro-ph",
    doi = "10.1088/0004-637X/690/1/L14",
    journal = "Astrophys. J. Lett.",
    volume = "690",
    pages = "L14--L17",
    year = "2009"
}

@article{Takami:2011nn,
    author = "Takami, Hajime and Murase, Kohta",
    title = "{The Role of Structured Magnetic Fields on Constraining Properties of Transient Sources of Ultra-high-energy Cosmic Rays}",
    eprint = "1110.3245",
    archivePrefix = "arXiv",
    primaryClass = "astro-ph.HE",
    doi = "10.1088/0004-637X/748/1/9",
    journal = "Astrophys. J.",
    volume = "748",
    pages = "9",
    year = "2012"
}

@article{Ehlert:2022jmy,
    author = "Ehlert, Domenik and Oikonomou, Foteini and Unger, Michael",
    title = "{Curious case of the maximum rigidity distribution of cosmic-ray accelerators}",
    eprint = "2207.10691",
    archivePrefix = "arXiv",
    primaryClass = "astro-ph.HE",
    doi = "10.1103/PhysRevD.107.103045",
    journal = "Phys. Rev. D",
    volume = "107",
    number = "10",
    pages = "103045",
    year = "2023"
}

@article{Coleman:2022abf,
    author = "Coleman, A. and others",
    title = "{Ultra high energy cosmic rays The intersection of the Cosmic and Energy Frontiers}",
    eprint = "2205.05845",
    archivePrefix = "arXiv",
    primaryClass = "astro-ph.HE",
    reportNumber = "FERMILAB-PUB-22-413-PPD",
    doi = "10.1016/j.astropartphys.2023.102819",
    journal = "Astropart. Phys.",
    volume = "149",
    pages = "102819",
    year = "2023"
}

@article{PierreAuger:2019ens,
    author = "Aab, A. and others",
    collaboration = "Pierre Auger",
    title = "{Probing the origin of ultra-high-energy cosmic rays with neutrinos in the EeV energy range using the Pierre Auger Observatory}",
    eprint = "1906.07422",
    archivePrefix = "arXiv",
    primaryClass = "astro-ph.HE",
    reportNumber = "FERMILAB-PUB-19-280-ND-PPD-TD",
    doi = "10.1088/1475-7516/2019/10/022",
    journal = "JCAP",
    volume = "10",
    pages = "022",
    year = "2019"
}

@article{PierreAuger:2013waq,
    author = "Abreu, Pedro and others",
    collaboration = "Pierre Auger",
    title = "{Bounds on the Density of Sources of Ultra-High Energy Cosmic Rays from the Pierre Auger Observatory}",
    eprint = "1305.1576",
    archivePrefix = "arXiv",
    primaryClass = "astro-ph.HE",
    reportNumber = "FERMILAB-PUB-13-148-AD-AE-CD-TD",
    doi = "10.1088/1475-7516/2013/05/009",
    journal = "JCAP",
    volume = "05",
    pages = "009",
    year = "2013"
}

@article{AlvesBatista:2019tlv,
    author = "Alves Batista, Rafael and others",
    title = "{Open Questions in Cosmic-Ray Research at Ultrahigh Energies}",
    eprint = "1903.06714",
    archivePrefix = "arXiv",
    primaryClass = "astro-ph.HE",
    doi = "10.3389/fspas.2019.00023",
    journal = "Front. Astron. Space Sci.",
    volume = "6",
    pages = "23",
    year = "2019"
}

@article{PierreAuger:2024hlp,
    author = "Abdul Halim, A. and others",
    collaboration = "Pierre Auger",
    title = "{Impact of the magnetic horizon on the interpretation of the Pierre Auger Observatory spectrum and composition data}",
    eprint = "2404.03533",
    archivePrefix = "arXiv",
    primaryClass = "astro-ph.HE",
    reportNumber = "FERMILAB-PUB-24-0144-CSAID-PPD-TD-V",
    doi = "10.1088/1475-7516/2024/07/094",
    journal = "JCAP",
    volume = "07",
    pages = "094",
    year = "2024"
}

@article{Jansson:2012pc,
    author = "Jansson, Ronnie and Farrar, Glennys R.",
    title = "{A New Model of the Galactic Magnetic Field}",
    eprint = "1204.3662",
    archivePrefix = "arXiv",
    primaryClass = "astro-ph.GA",
    doi = "10.1088/0004-637X/757/1/14",
    journal = "Astrophys. J.",
    volume = "757",
    pages = "14",
    year = "2012"
}

@article{Unger:2023lob,
    author = "Unger, Michael and Farrar, Glennys R.",
    title = "{The Coherent Magnetic Field of the Milky Way}",
    eprint = "2311.12120",
    archivePrefix = "arXiv",
    primaryClass = "astro-ph.GA",
    doi = "10.3847/1538-4357/ad4a54",
    journal = "Astrophys. J.",
    volume = "970",
    number = "1",
    pages = "95",
    year = "2024"
}

@article{Bister:2024ocm,
    author = "Bister, Teresa and Farrar, Glennys R. and Unger, Michael",
    title = "{The Large-scale Anisotropy and Flux (de)magnification of Ultrahigh-energy Cosmic Rays in the Galactic Magnetic Field}",
    eprint = "2408.00614",
    archivePrefix = "arXiv",
    primaryClass = "astro-ph.HE",
    doi = "10.3847/2041-8213/ad856f",
    journal = "Astrophys. J. Lett.",
    volume = "975",
    number = "1",
    pages = "L21",
    year = "2024"
}

@article{CRPropa:2022ovg,
    author = "Alves Batista, Rafael and others",
    collaboration = "CRPropa",
    title = "{CRPropa 3.2 {\textemdash} an advanced framework for high-energy particle propagation in extragalactic and galactic spaces}",
    eprint = "2208.00107",
    archivePrefix = "arXiv",
    primaryClass = "astro-ph.HE",
    doi = "10.1088/1475-7516/2022/09/035",
    journal = "JCAP",
    volume = "09",
    pages = "035",
    year = "2022"
}

@article{PierreAuger:2024fgl,
    author = "Halim, A. Abdul and others",
    collaboration = "Pierre Auger",
    title = "{Large-scale Cosmic-ray Anisotropies with 19 yr of Data from the Pierre Auger Observatory}",
    eprint = "2408.05292",
    archivePrefix = "arXiv",
    primaryClass = "astro-ph.HE",
    reportNumber = "FERMILAB-PUB-24-1012-V",
    doi = "10.3847/1538-4357/ad843b",
    journal = "Astrophys. J.",
    volume = "976",
    number = "1",
    pages = "48",
    year = "2024"
}

@article{Planck:2016gdp,
    author = "Adam, R. and others",
    collaboration = "Planck",
    title = "{Planck intermediate results.}: {XLII. Large-scale Galactic magnetic fields}",
    eprint = "1601.00546",
    archivePrefix = "arXiv",
    primaryClass = "astro-ph.GA",
    doi = "10.1051/0004-6361/201528033",
    journal = "Astron. Astrophys.",
    volume = "596",
    pages = "A103",
    year = "2016"
}

@article{AlvesBatista:2016vpy,
    author = {Alves Batista, Rafael and Dundovic, Andrej and Erdmann, Martin and Kampert, Karl-Heinz and Kuempel, Daniel and M\"uller, Gero and Sigl, Guenter and van Vliet, Arjen and Walz, David and Winchen, Tobias},
    title = "{CRPropa 3 - a Public Astrophysical Simulation Framework for Propagating Extraterrestrial Ultra-High Energy Particles}",
    eprint = "1603.07142",
    archivePrefix = "arXiv",
    primaryClass = "astro-ph.IM",
    doi = "10.1088/1475-7516/2016/05/038",
    journal = "JCAP",
    volume = "05",
    pages = "038",
    year = "2016"
}

@article{Greisen:1966jv,
    author = "Greisen, Kenneth",
    title = "{End to the cosmic ray spectrum?}",
    doi = "10.1103/PhysRevLett.16.748",
    journal = "Phys. Rev. Lett.",
    volume = "16",
    pages = "748--750",
    year = "1966"
}

@article{Marafico:2024qgh,
    author = "Marafico, Sullivan and Biteau, Jonathan and Condorelli, Antonio and Deligny, Olivier and Bregeon, Johan",
    title = "{Closing the Net on Transient Sources of Ultrahigh-energy Cosmic Rays}",
    eprint = "2405.17179",
    archivePrefix = "arXiv",
    primaryClass = "astro-ph.HE",
    doi = "10.3847/1538-4357/ad5a11",
    journal = "Astrophys. J.",
    volume = "972",
    number = "1",
    pages = "4",
    year = "2024"
}

@article{PierreAuger:2017pzq,
    author = "Aab, Alexander and others",
    collaboration = "Pierre Auger",
    title = "{Observation of a Large-scale Anisotropy in the Arrival Directions of Cosmic Rays above $8 \times 10^{18}$ eV}",
    eprint = "1709.07321",
    archivePrefix = "arXiv",
    primaryClass = "astro-ph.HE",
    reportNumber = "FERMILAB-PUB-17-354",
    doi = "10.1126/science.aan4338",
    journal = "Science",
    volume = "357",
    number = "6537",
    pages = "1266--1270",
    year = "2017"
}

@article{Kaapa:2022qqn,
    author = {K\"a\"ap\"a, Alex and Kampert, Karl-Heinz and Mayotte, Eric},
    title = "{Propagation of extragalactic cosmic rays in the Galactic magnetic field}",
    doi = "10.22323/1.398.0088",
    journal = "PoS",
    volume = "EPS-HEP2021",
    pages = "088",
    year = "2022"
}

@article{Eichmann:2022ias,
    author = {Eichmann, Bj\"orn and Kachelrie\ss{}, Michael and Oikonomou, Foteini},
    title = "{Explaining the UHECR spectrum, composition and large-scale anisotropies with radio galaxies}",
    eprint = "2202.11942",
    archivePrefix = "arXiv",
    primaryClass = "astro-ph.HE",
    doi = "10.1088/1475-7516/2022/07/006",
    journal = "JCAP",
    volume = "07",
    number = "07",
    pages = "006",
    year = "2022"
}

@article{Kaapa:2022tey,
    author = {K{\"a}{\"a}p{\"a}, Alex and Kampert, Karl-Heinz and Mayotte, Eric},
    title = "{The effects of the GMF on the transition from Galactic to extragalactic cosmic rays}",
    doi = "10.22323/1.395.0004",
    journal = "PoS",
    volume = "ICRC2021",
    pages = "004",
    year = "2022"
}

@article{Farrar:2017lhm,
    author = "Farrar, Glennys R. and Sutherland, Michael S.",
    title = "{Deflections of UHECRs in the Galactic magnetic field}",
    eprint = "1711.02730",
    archivePrefix = "arXiv",
    primaryClass = "astro-ph.HE",
    doi = "10.1088/1475-7516/2019/05/004",
    journal = "JCAP",
    volume = "05",
    pages = "004",
    year = "2019"
}

@article{Harari:2000az,
    author = "Harari, Diego and Mollerach, Silvia and Roulet, Esteban",
    title = "{Signatures of galactic magnetic lensing upon ultrahigh-energy cosmic rays}",
    eprint = "astro-ph/0001084",
    archivePrefix = "arXiv",
    doi = "10.1088/1126-6708/2000/02/035",
    journal = "JHEP",
    volume = "02",
    pages = "035",
    year = "2000"
}

@article{Farrar:2024zsm,
    author = "Farrar, Glennys R.",
    title = "{Binary Neutron Star Mergers as the Source of the Highest Energy Cosmic Rays}",
    eprint = "2405.12004",
    archivePrefix = "arXiv",
    primaryClass = "astro-ph.HE",
    doi = "10.1103/PhysRevLett.134.081003",
    journal = "Phys. Rev. Lett.",
    volume = "134",
    number = "8",
    pages = "081003",
    year = "2025"
}

@article{Giacinti:2011uj,
    author = "Giacinti, G. and Kachelriess, M. and Semikoz, D. V. and Sigl, G.",
    title = "{Ultrahigh Energy Nuclei in the Turbulent Galactic Magnetic Field}",
    eprint = "1104.1141",
    archivePrefix = "arXiv",
    primaryClass = "astro-ph.HE",
    doi = "10.1016/j.astropartphys.2011.07.006",
    journal = "Astropart. Phys.",
    volume = "35",
    pages = "192--200",
    year = "2011"
}

@article{Stanev:2000fb,
    author = "Stanev, Todor and Engel, Ralph and Mucke, Anita and Protheroe, Raymond J. and Rachen, Jorg P.",
    title = "{Propagation of ultrahigh-energy protons in the nearby universe}",
    eprint = "astro-ph/0003484",
    archivePrefix = "arXiv",
    doi = "10.1103/PhysRevD.62.093005",
    journal = "Phys. Rev. D",
    volume = "62",
    pages = "093005",
    year = "2000"
}

@article{PierreAuger:2025eun,
    author = "Abdul Halim, Adila and others",
    collaboration = "Pierre Auger",
    title = "{Energy Spectrum of Ultrahigh-Energy Cosmic Rays across Declinations -90{\textdegree} to +44.8{\textdegree} as Measured at the Pierre Auger Observatory}",
    eprint = "2506.11688",
    archivePrefix = "arXiv",
    primaryClass = "astro-ph.HE",
    doi = "10.1103/p4l5-hxlf",
    journal = "Phys. Rev. Lett.",
    volume = "135",
    number = "24",
    pages = "241002",
    year = "2025"
}

@article{PierreAuger:2020kuy,
    author = "Aab, Alexander and others",
    collaboration = "Pierre Auger",
    title = "{Features of the Energy Spectrum of Cosmic Rays above 2.5{\texttimes}10$^{18}$ eV Using the Pierre Auger Observatory}",
    eprint = "2008.06488",
    archivePrefix = "arXiv",
    primaryClass = "astro-ph.HE",
    reportNumber = "FERMILAB-PUB-20-431-E-TD",
    doi = "10.1103/PhysRevLett.125.121106",
    journal = "Phys. Rev. Lett.",
    volume = "125",
    number = "12",
    pages = "121106",
    year = "2020"
}

\newpage

\appendix

\section{The Justification of the Use of the Lambertian Distribution}
\label{app:lambert}
In this appendix, we show that injecting particles from a spherical shell with a Lambertian distribution of momenta is equivalent to injecting particles from many distant isotropically emitting sources. We also argue that this means that we can select any source by selecting particles from our kernel that arrived at the Galaxy from that specific direction.

\begin{figure}[htb]
    \centering
    \includegraphics[width=0.5\linewidth]{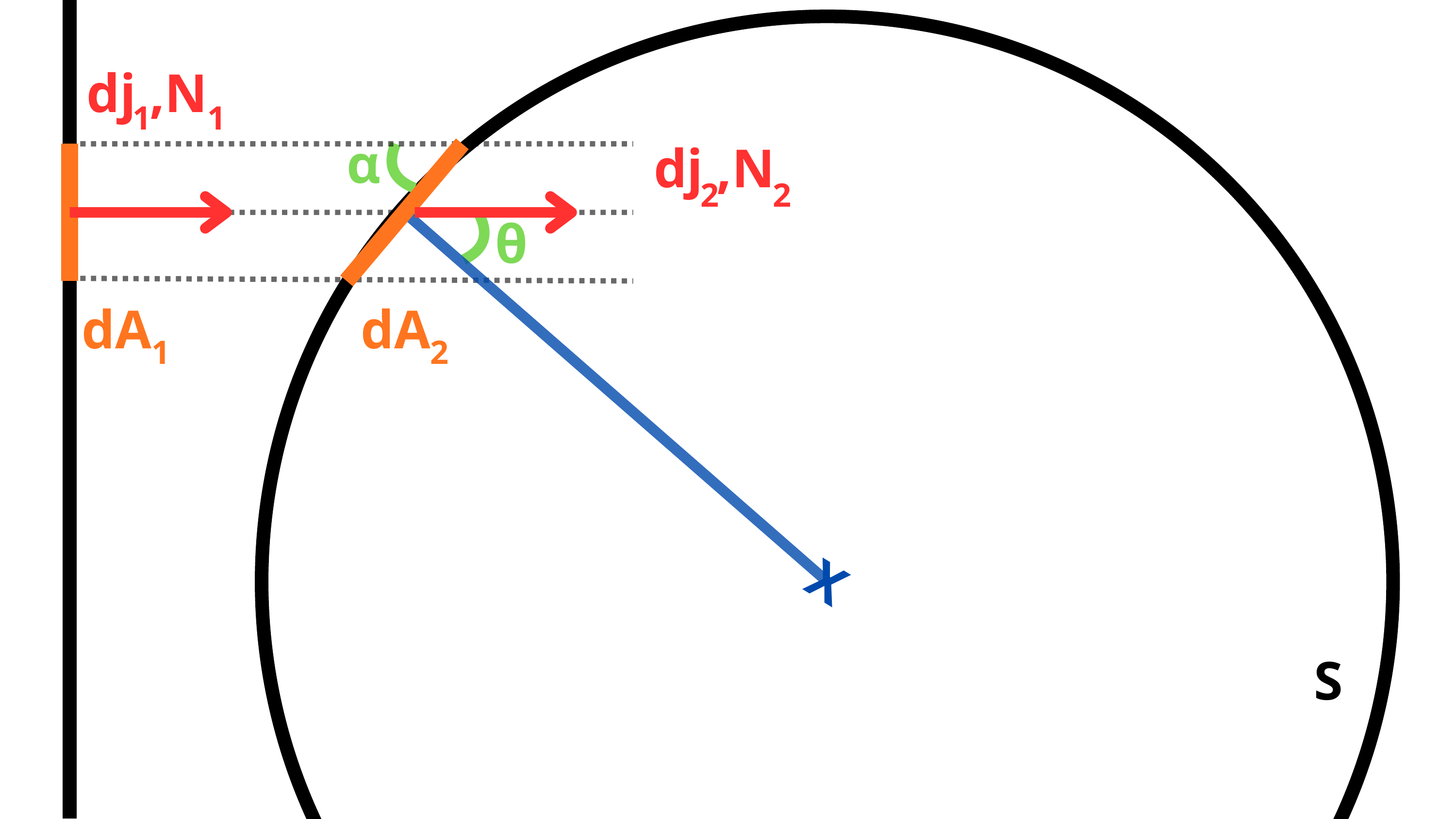}
    \caption{A sketch illustrating the geometrical argument behind using the Lambertian distribution. A plane wave of particles with scalar flux $j_1$ arriving to the spherical surface $s$.
    }
    \label{fig:lambert}
\end{figure}

The setup of this proof is illustrated in Fig.\,\ref{fig:lambert}. Suppose a source very distant from the Galaxy emits cosmic rays isotropically. Assuming ballistic propagation away from the source and neglecting pitch angle diffusion, the emission at the edge of the Galaxy can be approximated by a plane wave with a scalar flux, $dj_1 = N_1/(dA_1\,dt)$. This flux hits the spherical surface $s$, which, in our case, encapsulates the Galaxy at an angle $\alpha$. The corresponding flux is then $dj_2 = N_2/(dA_2\,dt)$. Since the number of particles is conserved, $N_1=N_2$, the surface elements are related by $dA_1=dA_2 \sin\alpha$. Therefore, $dj_2=dj_1 \sin\alpha$.

We realize that $\alpha=90^\circ - \theta$ in coordinates where the normal vector of $dA_2$ is $(\phi, \theta, z) = (0,0,1)$. Together with the identity $\sin(90^\circ - \theta)=\cos\theta$, this yields $dj_2=dj_1\cos\theta$, and we recover the Lambertian factor. Therefore, the probability of selecting a direction within a small range of $d\theta$ is $P(d\theta) \sim \cos(\theta) d\theta$.

Now, we zoom out of the 2D sketch in Fig.\,\ref{fig:lambert}, acknowledging that the plane wave can come from any direction. Equivalently, the momentum directions cover the entire hemisphere above the injection point, which has a solid angle of $\Omega=2\pi$. To select a momentum in the range of possible directions $d\Omega$, we have $P(d\Omega)=\frac{1}{2\pi}\cos\theta d\Omega$. After transforming the coordinates, this becomes $P(d\Omega)=\frac{1}{2\pi}\cos\theta \sin\theta\,d\theta d\phi$, and integrating over $\phi$, $P(d\Omega)= \cos\theta \sin\theta\,d\theta$. Thus, we obtain the expression to generate the vectors for injection from a specific direction (c.f.\ \cite{crpropadocs:lambert}).

We showed here that our complete kernel in fact represents the scenario of many distant, homogeneously distributed sources. It also follows that if only one initial direction is selected, the initial plane wave is recovered, and thus the source in the specified direction has been isolated.

\section{Other Magnetic Fields}
\label{app:otherB}
\begin{figure}[htb]
    \centering
    \includegraphics[width=0.95\linewidth]{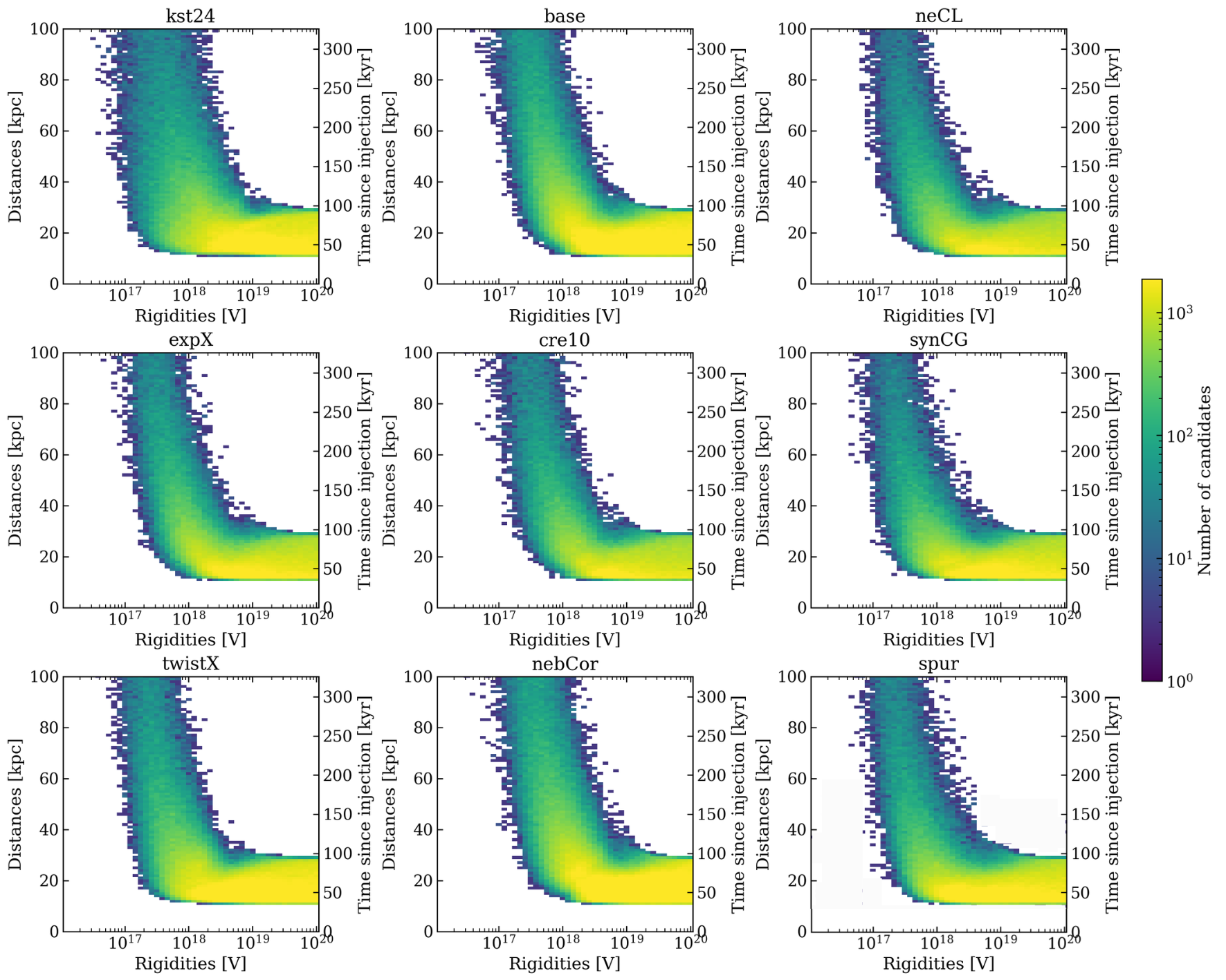}
    \caption{As in Fig.\,\ref{fig:rig_vs_dist}, however, using other models of the GMF, corresponding to figure headers. Each 2D histogram comprises of $10^{5}-10^{6}$ particles. 
    }
    \label{fig:otherKernels}
\end{figure}

\begin{figure}[htb]
    \centering
    \includegraphics[width=0.95\linewidth]{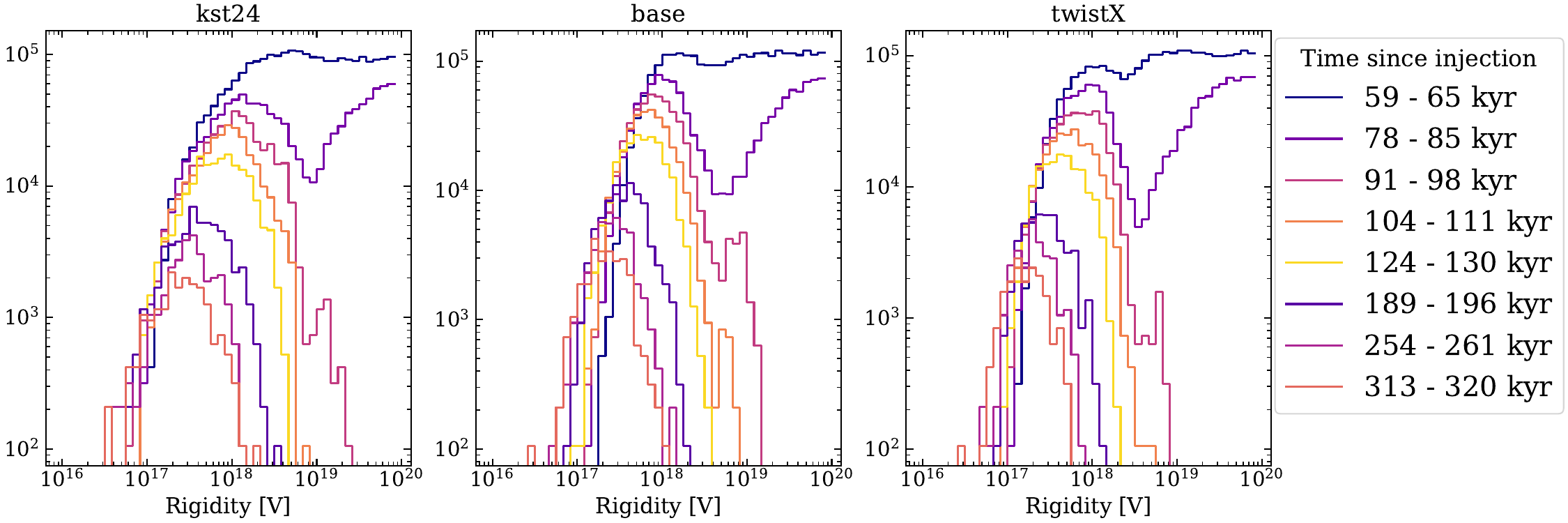}
    \caption{Vertical slices through selected kernels in  Fig.\,\ref{fig:otherKernels} detail how the GMF influences an originally flat ($R^{-1}$) spectrum, including all directions for generality. The cutoff develops after $\sim 100$\,kyr at rigidities around the $10^{19}$\,V and then moves to $\approx 10^{18}$\,V at later times. 
    }
    \label{fig:otherSlices}
\end{figure}

To determine how sensitive this framework is to different realizations of the fields' strength and geometry, we tested other GMF models. We ran simulations for eight models in the UF23 package and KST24, each time incorporating a random field from JF12. The results are remarkably similar to each other, as shown in Figs.\,\ref{fig:otherKernels} and \ref{fig:otherSlices}. This suggests that the galactic response depends more on the overall strength and geometry of the GMF than on its specific details.

There are slight differences between the models on where the delays are triggered, and how wide the distribution of rigidities is. Both depend on the overall strength of the field; however, the latter is also heavily influenced by a comparatively lower statistics. Note that the kernels in Fig.\,\ref{fig:otherKernels} are composed each of about 1000 times fewer particles than that in Fig.\,\ref{fig:rig_vs_dist}). The qualitative differences and their origin are a subjects of future work.

\section{Temporal Dependence of Correlations between Initial Positions and Directions}

\begin{figure}[tbh]
    \centering
    \includegraphics[width=.9\linewidth]{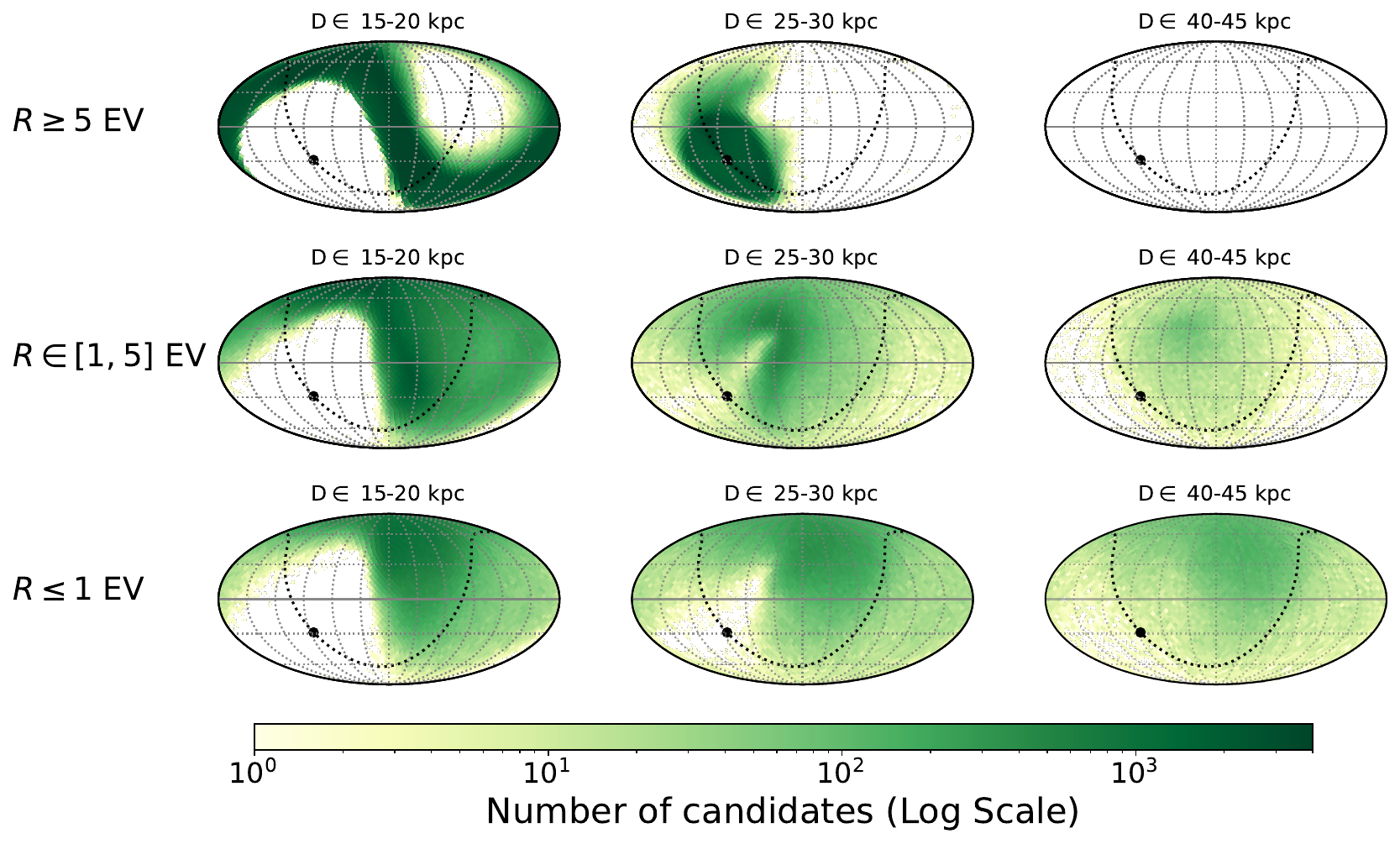}
    \caption{Distributions of the initial directions of  protons recorded at the observer sphere after propagation through the Galaxy. The skymaps grouped in three ranges of rigidity and propagation distance in the Galaxy.}
    \label{fig:initial_directions}
\end{figure}

Figure~\ref{fig:initial_directions} presents the initial directions of UHECRs recorded in three rigidity and three propagation distance ranges. The skymaps for $R>5$\,EV, where the effect of the GMF is minimized, demonstrate the expected correlation between initial direction and propagation distance. For intermediate rigidities, $R \in [1,5]$\,EV, such a correlation is notably disrupted as illustrated by the wider areas of the skymaps included with increased propagation distance. Although these maps assume a Lambertian initial distribution, the effects of the GMF generally apply: in the observed range of UHECR rigidities and below, both the direction and position at the edge of the Galaxy are important. Thus, the correlation between initial and observed directions cannot be established when considering the age of the UHECR in the Galaxy.

\begin{figure}[tbh]
    \centering
    \includegraphics[width=.9\linewidth]{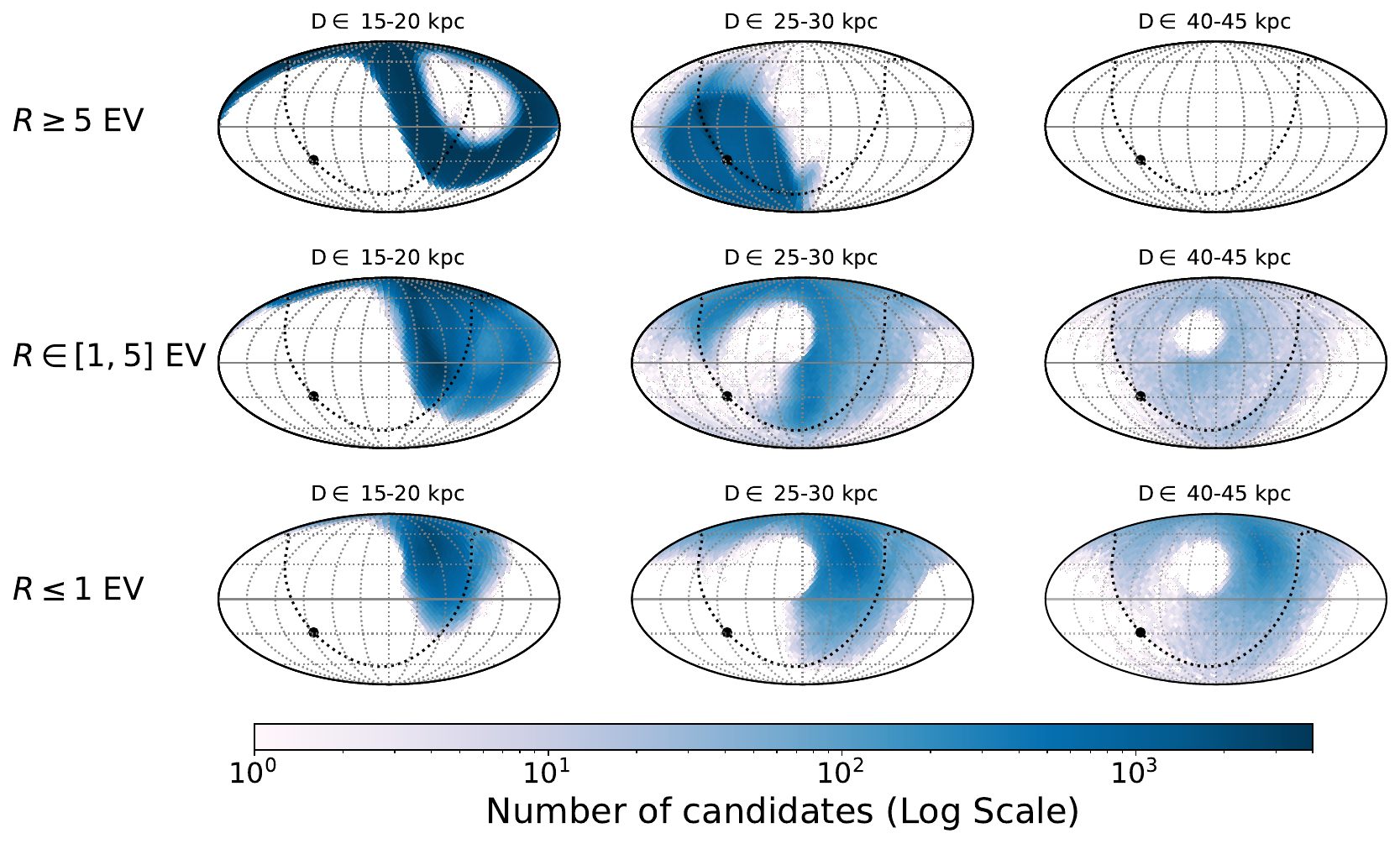}
    \caption{Same as Fig.\,\ref{fig:initial_directions}, but for the initial positions at the edge of the Galaxy.
    }
    \label{fig:initial_positions}
\end{figure}

Taking into account the distributions of the initial positions of the observed UHECRs, as shown in Fig.\,\ref{fig:initial_positions}, clarifies this. At the highest rigidities, the skymaps of positions are very similar to the skymaps of directions due to the aforementioned geometrical correlation. This correlation is approximately preserved for UHECRs with short propagation distances. However, it becomes increasingly compromised as the age of the cosmic rays increases. Ultimately, virtually all cosmic rays incident within a certain latitude band near the Galactic equator can be observed. However, the large deflection angles impede the establishment of a correlation between the initial and observed directions.

\section{Galactic Response Function to Burst-Like Injections from Different Directions}
\label{sec:directional-injection}

Section \ref{sec:MultipleInjections} discussed the Galactic response to multiple injections. In this Appendix, we demonstrate that the rigidity-dependent cutoff observed in the cosmic ray spectrum over an extended period of time is not only present for an isotropic injection (\textit{c.f.} Fig.\,\ref{fig:temporal_distribution}, left) but also a generic feature for cosmic rays injected into the Galaxy from any direction.

\begin{figure}[tbh]
    \centering
    \includegraphics[width=\linewidth]{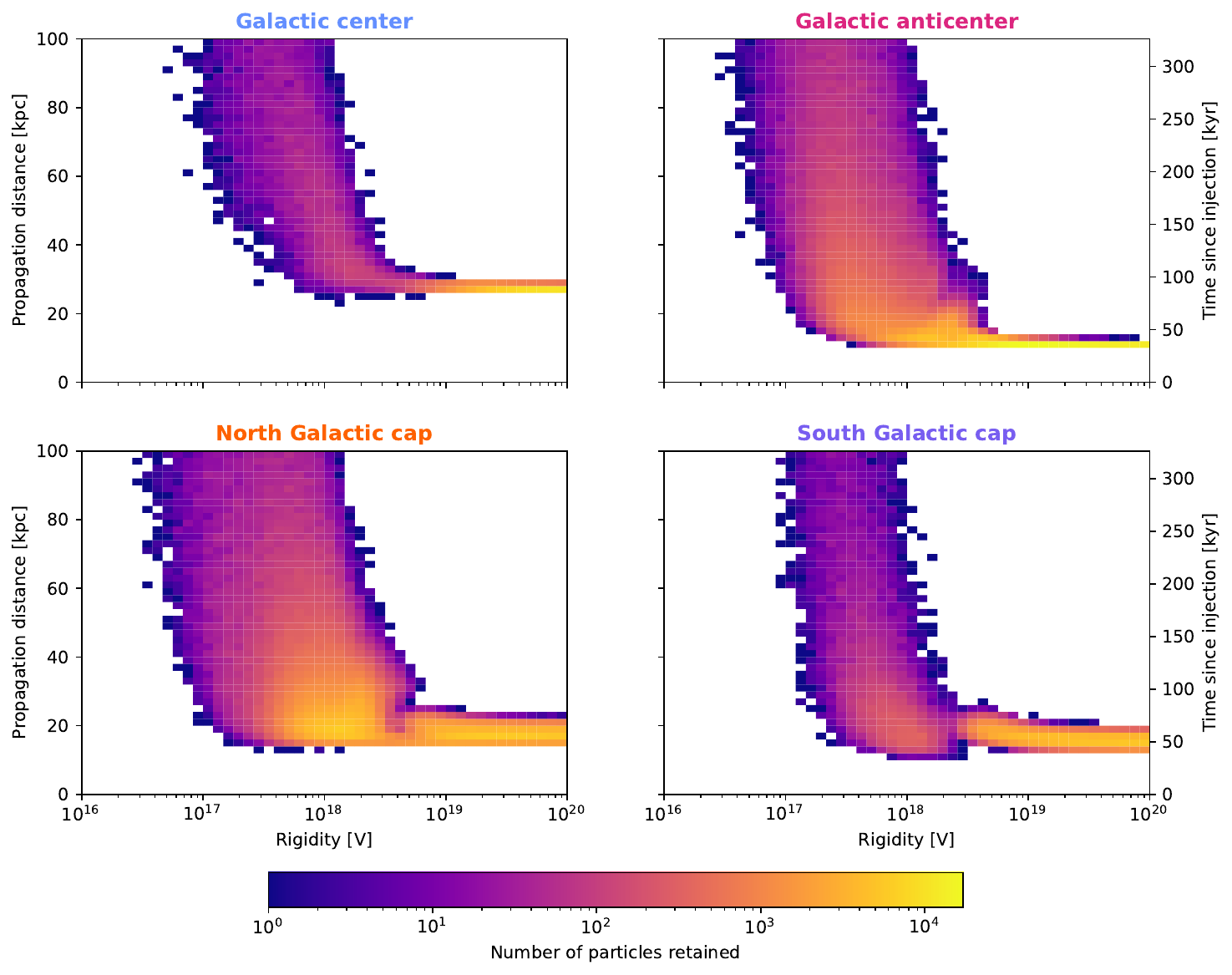}
    \caption{Same as Fig.\,\ref{fig:rig_vs_dist}, but selected for cosmic rays arriving at the edge of the Galaxy from four distinct directions: from behind the Galactic center, from the Galactic anticenter, and from the north and south Galactic cap regions.
    }
    \label{fig:directional-kernels}
\end{figure}

The response kernel shown in Fig.\,\ref{fig:rig_vs_dist} can be used to easily visualize this process for an instantaneous injection of cosmic rays from specific directions. Technically, this is achieved by selecting cosmic rays that entered the injection sphere from specific regions of the sky. For illustration, four directions with an angular opening of approximately $18^\circ$ each were chosen. These directions correspond to cosmic rays arriving at the north and south Galactic poles, from behind the Galactic center, and from the Galactic anticenter. The four response kernels are shown in Fig.\,\ref{fig:directional-kernels}. They all exhibit similar qualitative features: cosmic rays with $R\gtrsim 6\cdot 10^{18}$\,EV can only be observed within a short time frame after injection, whereas those with lower rigidities spend 300\,kyr or more in the Galaxy before arriving at Earth. During this period, the mean rigidity and high-rigidity cutoff decrease over time. However, there are also distinct differences related to the position of the observer sphere within the Galaxy and the characteristics of the GMF through which the cosmic rays must propagate before reaching Earth. Interestingly, these differences are more prominent for cosmic rays with low than those with high rigidities. The lowest integrated magnetic field strength along the cosmic ray path to Earth, $\int B(s)ds$, is expected for cosmic rays arriving from the Galactic anticenter and the north Galactic cap. From these directions, also cosmic rays with lower rigidity can reach Earth within reasonably short propagation distances, so that the corresponding kernels extend to lower rigidities. Conversely, low-rigidity cosmic rays passing through the Galactic center region or through the southern GMF, which is known to be stronger than that of the northern Galactic hemisphere, experience much longer path lengths, before reaching Earth. High rigidity cosmic rays are affected in the same way, but to a much lesser extent so that the position of the high energy cutoff is less influenced by the direction from which the cosmic rays arrive. 

\end{document}